\documentclass[twocolumn]{aastex701}

\shorttitle{LRDs at Cosmic Noon}
\shortauthors{D. J. Lyon. et al.}

\received{Month 1, 20XX}
\revised{Month 1, 20XX}
\accepted{Month 1, 20XX}

\submitjournal{ApJ}

\graphicspath{{./}{Figures/}}

\usepackage{lipsum}
\usepackage{booktabs}
\usepackage{amsmath}
\usepackage{enumitem}
\usepackage{iftex}
\ifXeTeX
  \usepackage{xeCJK}
  \newcommand{\cjkname}[1]{#1}
\else
  \usepackage{CJKutf8}
  \newcommand{\cjkname}[1]{\begin{CJK*}{UTF8}{gbsn}#1\end{CJK*}}
\fi

\begin{document}

\title{Little Red Dots at Cosmic Noon: Spectroscopic Identification and Broad-Line Properties}
%

\correspondingauthor{Daniel Lyon (djlyon@swin.edu.au)}
\author[orcid=0000-0002-7847-1107]{Daniel J. Lyon}
\affiliation{Centre for Astrophysics and Supercomputing, Swinburne University of Technology, Melbourne, VIC 3122, Australia}
\email[show]{djlyon@swin.edu.au}  

\author[0000-0002-0463-9528]{Yilun Ma (\cjkname{马逸伦})}
\affiliation{Department of Astrophysical Sciences, Princeton University, Princeton, NJ 08544, USA}
\email{yilun@princeton.edu}

\author[0000-0002-2057-5376]{Ivo Labbe}
\affiliation{Centre for Astrophysics and Supercomputing, Swinburne University of Technology, Melbourne, VIC 3122, Australia}
\email{ilabbe@swin.edu.au}

\author[0000-0001-5586-6950]{Alberto Torralba}
\affiliation{Institute of Science and Technology Austria (ISTA), Am Campus 1, 3400 Klosterneuburg, Austria}
\email{alberto.torralba@ista.ac.at}

\author[0000-0002-5612-3427]{Jenny E. Greene}
\affiliation{Department of Astrophysical Sciences, Princeton University, Princeton, NJ 08544, USA}
\email{jgreene@astro.princeton.edu}

\author[0000-0003-2871-127X]{Jorryt Matthee} 
\affiliation{Institute of Science and Technology Austria (ISTA), Am Campus 1, 3400 Klosterneuburg, Austria}
\email{jorryt.matthee@ist.ac.at} 

\author[orcid=0000-0002-3254-9044]{Karl Glazebrook} 
\affiliation{Centre for Astrophysics and Supercomputing, Swinburne University of Technology, Melbourne, VIC 3122, Australia}
\email{kglazebrook@swin.edu.au}  

\author[0000-0003-4700-663X]{Andy D. Goulding} 
\affiliation{Department of Astrophysical Sciences, Princeton University, Princeton, NJ 08544, USA}
\email{goulding@astro.princeton.edu}

\author[0000-0003-4075-7393]{David J. Setton}\thanks{NHFP Hubble Fellow}
\affiliation{The William H. Miller III Department of Physics and Astronomy, 3400 N. Charles Street, Baltimore, MD 21218}
\email{davidsetton@jhu.edu}

\author[orcid=0000-0003-2804-0648]{Themiya Nanayakkara} 
\affiliation{Sydney Institute for Astronomy, School of Physics A28, The University of Sydney, NSW 2006, Australia.}
\email{themiya.nanayakkara@sydney.edu.au}

\author[0000-0002-8053-8040]{Marianna Annunziatella}
\affiliation{INAF-IASF Milano, via A. Corti 12, I-20133 Milano, Italy}
\email{mannunziatella@cab.inta-csic.es}

\author[0009-0008-9877-5512]{Aayush Desai}
\affiliation{Institute of Science and Technology Austria (ISTA), Am Campus 1, 3400 Klosterneuburg, Austria}
\email{aayush.desai@ist.ac.at}

\author[0000-0003-1344-9475]{Eiichi Egami} 
\affiliation{Steward Observatory, University of Arizona, 933 N Cherry Avenue, Tucson, AZ 85721, USA}
\email{egami@arizona.edu}

\author[0000-0003-3310-0131]{Xiaohui Fan}
\affiliation{Steward Observatory, University of Arizona, 933 N Cherry Avenue, Tucson, AZ 85721, USA}
\email{xiaohuidominicfan@gmail.com}

\author[0000-0002-5588-9156]{Vasily Kokorev}
\affiliation{Department of Astronomy, The University of Texas at Austin, Austin, TX 78712, USA}
\email{vkokorev@utexas.edu}

\author[0000-0001-9592-4190]{Jiaxuan Li (\cjkname{李嘉轩})}
\email{jiaxuanl@stanford.edu}
\affiliation{Kavli Institute for Particle Astrophysics and Cosmology, Stanford University, Stanford, CA 94305, USA}

\author[0000-0001-6052-4234]{Xiaojing Lin}
\affiliation{Department of Astronomy, Tsinghua University, Beijing 100084, People's Republic of China}
\email{xiaojinglin.astro@gmail.com}

\author[0000-0001-9002-3502]{Danilo Marchesini}
\affiliation{Department of Physics and Astronomy, Tufts University, Medford, MA 02155, USA}
\email{danilo.marchesini@tufts.edu}

\author[0000-0002-6265-2675]{Luke Robbins}
\affiliation{Department of Physics and Astronomy, Tufts University, Medford, MA 02155, USA}
\email{andrew.robbins@tufts.edu}

\author[0000-0002-1917-1200]{Anna Sajina}
\affiliation{Department of Physics and Astronomy, Tufts University, Medford, MA 02155, USA}
\email{annie.sajina@gmail.com}

\author[0000-0002-7712-7857]{Marcin Sawicki}
\affiliation{Institute for Computational Astrophysics and Department of Astronomy and Physics, Saint Mary's University, 923 Robie Street, Halifax, Nova Scotia, B3H 3C3, Canada}
\affiliation{NRC-Herzberg, 5071 West Saanich Road, Victoria, British Columbia V9E 2E7, Canada}
\email{marcin.sawicki@smu.ca}

\begin{abstract}

We present Keck/NIRES, VLT/X-shooter, and Magellan/FIRE spectroscopic follow-up observations of photometrically selected Little Red Dot (LRD) candidates targeting H$\alpha$ at $2<z<3$. The apparent decline of LRDs at cosmic noon raises the question whether they retain the spectra of the high-redshift population. From a parent sample of 266 unresolved sources with V-shaped broadband SEDs, identified in 19.9\,deg$^2$ of the Hyper Suprime-Cam Subaru Strategic Program Deep/UltraDeep fields, we observe 24 and confirm broad H$\alpha$ in 6. The remaining 18 candidates show no broad H$\alpha$. Four have Ly$\alpha$ redshifts, which place H$\alpha$ outside our coverage for two. The other 14 have only photometric redshifts, so their H$\alpha$ coverage is uncertain; they may have incorrect photometric redshifts, broad H$\alpha$ fainter than $L_{\mathrm{H}\alpha}\approx3\times10^{42}$ erg s$^{-1}$, or not be LRDs. All six broad H$\alpha$ lines are best fit by exponential profiles, with FWHM $\approx 950$--$2150$ km s$^{-1}$. Three show Balmer absorption, all show Balmer decrements above Case B ($H\alpha/H\beta \approx 8$--$34$, though H$\beta$ is not detected), and four have V-shaped SEDs consistent with JWST-selected LRD samples. Archival \textit{JWST}/NIRCam imaging and NIRSpec spectroscopy independently confirm one source as an LRD. One of the six is X-ray detected at a luminosity typical of ordinary AGN and is better described as a broad-line AGN. Only one of seven dust-reddened F2M quasars and two of 38 heavily reddened quasars at $2<z<3$ meet typical JWST V-shape criteria. Four of the six sources with broad H$\alpha$ ($67$\%) resemble JWST-selected LRDs, showing that ground-based photometric selection and spectroscopy can identify LRDs.

\end{abstract}

\keywords{\uat{Active galactic nuclei}{16} --- \uat{Supermassive black holes}{1663} --- \uat{High-redshift galaxies}{734} --- \uat{Galaxy evolution}{594} --- \uat{Quasars}{1319} --- \uat{Galaxy spectroscopy}{2171}}
\section{Introduction}
\label{Sec: Introduction}

The \textit{James Webb Space Telescope} (\textit{JWST}) has revealed a population of compact, red objects that went previously undetected, now known as \textit{Little Red Dots} (LRDs; \citealp{labbe_population_2023, matthee_little_2024, kocevski_rise_2025}). They are often unresolved in JWST imaging ($r_e \approx 50$--$300$ pc: \citealp{kokorev_census_2024, furtak_high_2024}) and display a characteristic ``V''-shaped spectral energy distribution (SED) which is blue in the rest-frame UV and red in the rest-frame optical \citep[e.g.,][]{kokorev_census_2024}. They are found predominantly at $4 < z < 8$ \citep{akins_cosmos-web_2024, kocevski_rise_2025} where they outnumber comparably luminous UV-selected quasars by at least 1\,dex \citep[e.g.,][]{kokorev_census_2024} and may account for at least a few percent of galaxies at comparable redshift and stellar mass \citep{perez-gonzalez_what_2024}.

Most LRDs show broad Balmer emission \citep{greene_uncover_2024, degraaff_little_2025, scholtz_little_2026}, with widths and luminosities exceeding what star formation can plausibly produce \citep{matthee_little_2024, inayoshi_extremely_2025, de_graaff_remarkable_2025}, suggesting active galactic nucleus (AGN) activity. However, LRDs are undetected in the deepest \textit{Chandra} exposures \citep{yue_stacking_2024, maiolino_jwst_2025, sacchi_chandra_2025}, and show little to no variability \citep{kokubo_challenging_2024, zhou_variability_2025, liu_how_2026}. Their Balmer lines frequently show absorption, with non-Gaussian profiles (Lorentzian, exponential) and wings \citep{juodzbalis_jades_2024, kokorev_deepest_2025, scholtz_little_2026, madau_wings_2026, matthee_engine_2026}, alongside Balmer decrements far above expectation from Case B recombination \citep{deugenio_irony_2025, inayoshi_extremely_2025, torralba_warm_2025} that cannot be explained by standard dust extinction, although even that has been challenged \citep{nikopoulos_evidence_2026}. Together these signatures have been interpreted as an accretion source embedded in dense gas \citep{inayoshi_extremely_2025, de_graaff_remarkable_2025, naidu_black_2025}, a configuration rare among typical AGN. Whether that gas fully encloses the source, or is seen only along high-inclination sightlines, remains open \citep{madau_little_2026, barrufet_orientation_2026}.

Although they are extremely abundant at high redshift, LRDs seem to disappear towards later cosmic epochs \citep{kocevski_rise_2025}. Yet, how fast this abundance decline occurs is similarly contested \citep{euclidcollaboration_euclid_2025, kapoor_little_2026, loiacono_no_2026, ma_counting_2025}. The redshift evolution of LRDs can give clues to the conditions under which LRDs form, and it needs to be pinned down to identify physical drivers that distinguish LRDs from typical AGN. Cosmic noon provides a useful test of the connection between LRDs and the broader accreting black hole population. Ordinary quasar activity is abundant at this epoch, whereas LRDs are quite rare. Determining whether cosmic-noon candidates exhibit the characteristic spectral properties of \textit{JWST}-selected LRDs will reveal whether the physical conditions giving rise to these properties persist at later cosmic epochs. These sources are also accessible to ground-based facilities: at $2 < z < 3$, H$\alpha$ falls in the near-infrared, and wide-field imaging samples the V-shaped SED over areas far larger than current \textit{JWST} surveys. Follow-up ground-based spectroscopy can reach greater resolution than JWST offers, resolving emission-line profiles and recovering Balmer absorption that is readily missed at low resolution \citep{deugenio_jades_2025, torralba_black_2026}. Comparisons with dust-reddened quasars, whose continuum colors sometimes overlap those of LRDs \citep{glikman_first2mass_2012, stepney_rest_2026}, can assess contamination and test whether the populations share similar spectral properties and physical conditions.

Photometric searches have begun to identify candidates at $z < 4$ \citep{ma_counting_2025, euclidcollaboration_euclid_2025} with a small number of individual analogues spectroscopically confirmed \citep{ji_lord_2025, juodzbalis_jades_2024, fu_discovery_2025}, and LRD-like analogues have been reported as low as $z < 0.4$ \citep{lin_discovery_2025, ding_discovery_2026}. In contrast to the decline seen with \textit{JWST}, \cite{euclidcollaboration_euclid_2025} find the number density of their candidates increasing down to $z \approx 1.5$--$2.5$, emphasizing the need for spectroscopic confirmation of LRD candidates at cosmic noon. Ground- and space-based color selection at these redshifts has rarely been followed up spectroscopically.

In this paper we test whether ground-based selection can reliably recover LRDs at cosmic noon with the same characteristics as JWST samples with near-infrared spectroscopy of LRD candidates at cosmic noon selected photometrically by Y. Ma et al. (2026c, submitted) from Hyper Suprime-Cam imaging and observed with Keck/NIRES, VLT/X-shooter, and Magellan/FIRE. We ask four questions: (i) how efficiently does ground-based selection at $2 < z < 3$ recover genuine LRDs? (ii) do the signatures that characterize the high-redshift population persist at cosmic noon, namely a V-shaped continuum, broad H$\alpha$ with non-Gaussian wings, Balmer absorption, elevated Balmer decrements, and X-ray weakness? (iii) do these properties separate LRDs from the reddened quasars that share their colors? and (iv) how do their broad-line widths and luminosities compare with those of JWST LRDs, and what do they imply for black hole masses and accretion rates?

Section \ref{Sec: Data} describes the parent photometry, target selection, observations, and data reduction. Section \ref{Sec: Results} presents the spectra, line modeling, continuum slopes, Balmer decrements, and X-ray observations. Section \ref{Sec: Discussion} discusses these results in context with the literature, and Section \ref{Sec: Conclusion} summarizes our findings. Throughout this paper we adopt a cosmology of $H_0 = 70$ km $\mathrm{s^{-1}\ Mpc^{-1}}$, $\Omega_m=0.3$, and $\Omega_\Lambda=0.7$.
\section{Data} \label{Sec: Data}

In this section, we describe the various data used in this work. These include multiwavelength imaging from various wide-field surveys, our ground-based spectroscopy (from Keck, VLT, and Magellan), archival JWST spectroscopy, and archival Chandra observations.

\subsection{Multiwavelength Imaging Surveys} \label{Sec: Data Photometric}

The parent photometric catalog from which we select our sample is the so-called $u$-to-IRAC catalog. It contains deep $grizy$ imaging from the Deep/UltraDeep fields of the Hyper Suprime-Cam Subaru Strategic Program (HSC-SSP) that reaches a $5\sigma$ point-source detection limit of $i = 26.9$ \citep{aihara_hyper_2018, aihara_third_2022}. The size of the point spread function in the $i$-band is $0.66''$ \citep{aihara_third_2022}. The optical photometry is supplemented by $u/u^*$-band imaging from the CFHT Large Area $U$-band Survey \citep[CLAUDS;][]{sawicki_cfht_2019} and near-infrared $JHK/K_{s}$ imaging\footnote{For simplicity, we hereafter use $K$ to refer to both the WFCAM/$K$-band and VIRCAM/$K_\mathrm{s}$-band, unless specified otherwise.} from the Ultra Deep Survey with the VISTA Telescope (UltraVISTA; \citealt{mccracken_ultravista_2012}), the VISTA Deep Extragalactic Observations Survey (VIDEO; \citealt{jarvis_vista_2013}), the Deep UKIRT Near-infrared Steward Survey (DUNES$^2$; \citealt{aihara_third_2022}; E. Egami et al., in prep.), the
Deep eXtragalactic Survey and Ultra-Deep Survey of the UKIRT Infrared Deep Sky Survey (UKIDSS/DXS and UKIDSS/UDS; \citealt{Lawrence2007ukidss}), and the DeepCos survey (PI: Y.-T. Lin; \citealt{aihara_third_2022}). We also include \textit{Spitzer}/IRAC imaging at 3.6 and 4.5 $\mu$m assembled by \cite{Lacy2021deepdrill} and the Spitzer Coverage of the HSC-Deep with IRAC for Z-studies (SHIRAZ; \citealt{Annunziatella2023shiraz}). Source detection is performed using a modified HSC pipeline \citep{bosch_hyper_2018, Bosch2019lsstpipe, Desprez2023clauds+hsc} for imaging from $u$/$u^*$ through $K/K_\mathrm{s}$ and a customized deblending technique for the IRAC bands \citep[e.g.,][]{Labbe2006, Marchesini2009smf}, followed by forced photometry across all bands. A detailed description of how this catalog is constructed and used for selection in our work can be found in \cite{ma_counting_2025} as well as our companion paper (Y. Ma et al. 2026c, submitted). 

\subsection{Selection} \label{Sec: Data Selection}

\begin{figure}[h!]
    \centering
    \includegraphics[width=\linewidth]{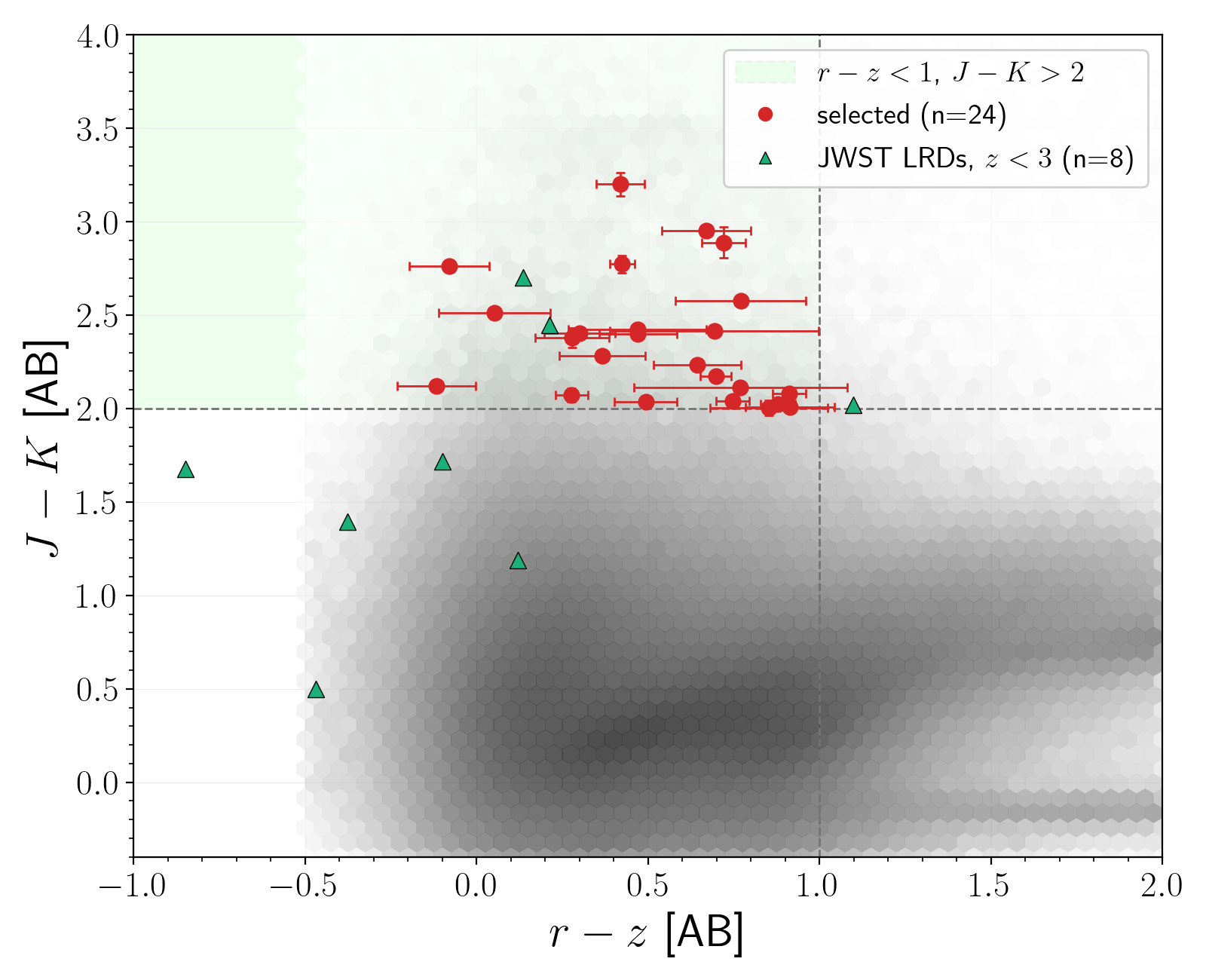}
    \caption{Color--color selection diagram. Candidates inside the box are selected with $r - z < 1$ \& $J - K > 2$. Grey hexagons show the parent catalog. Green triangles are the $z < 3$ JWST LRDs of Figure \ref{Fig: Beta Slopes} \citep{degraaff_little_2025}, with colors synthesized from their NIRSpec PRISM spectra.}
    \label{Fig: color-color}
\end{figure}

We select LRD candidates from this $u$-to-IRAC catalog by color rather than by fitting a photometric slope. The detailed selection is in Y. Ma et al. (2026c, submitted). At $2 < z < 4$ the rest-frame UV falls in the HSC $grizy$ bands and the rest-frame optical in the near-infrared. A V-shaped SED is then simply a blue $r - z$ paired with a red $J - K$ (Figure \ref{Fig: color-color}).

\begin{enumerate}
    \item $S/N>3$ from $r$-band to $K$-band;
    \item the object is consistent with a point source ($m_\mathrm{PSF}-m_\mathrm{CMODEL}<\Delta m_\mathrm{threshold}$; \citealt{bosch_hyper_2018}) in the $i$, $J$, $H$, and $K$ bands
    \item $J - K > 2$
    \item $r - z < 1$
    \item $y - J > -0.4$
    \item $-1.0 < K - [3.6] < 1.7$ (where Spitzer/IRAC is available)
\end{enumerate}

Signal-to-noise and point-source requirements remove extended and marginally detected objects, and color cuts in the near-infrared reject extremely red quasars \citep{hamann_extremely_2017} and extremely luminous hot dust-obscured galaxies \citep{tsai_most_2015}.

Our photometric candidates were drawn from the subset of this catalog at $1.7<z<3.7$ resulting in 266 objects in total. From these we ranked the candidates with a weighting system in which brighter objects are prioritized. See Y. Ma et al. (2026c, submitted) for details.  

\subsection{Keck/NIRES Spectroscopy} \label{Sec Data Spectroscopy}

\begin{figure*}[ht!]
    \centering
    \includegraphics[width=\linewidth]{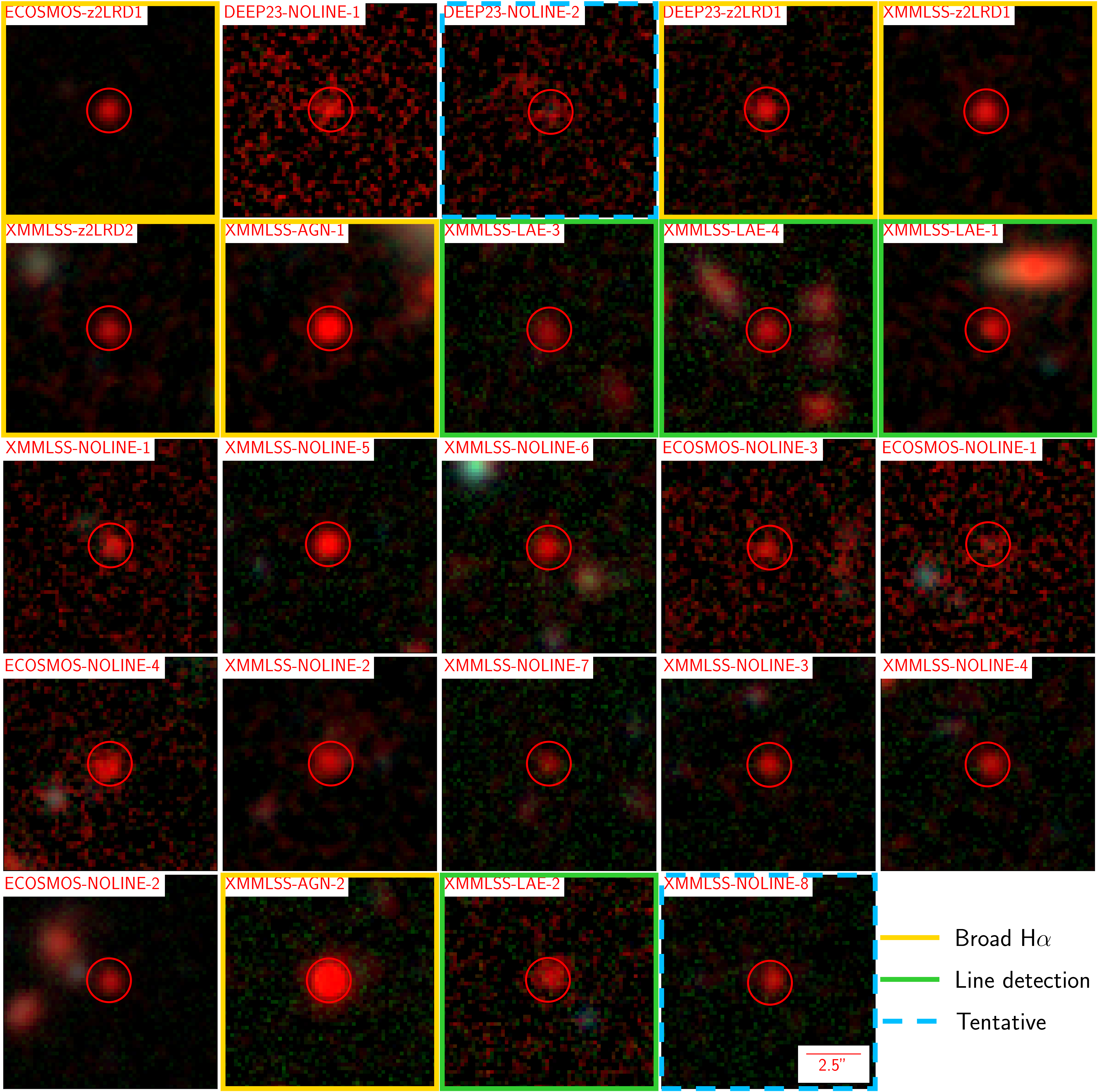}
    \caption{Mosaic view of targets observed with Keck/NIRES, VLT/X-shooter and Magellan/FIRE. Each image is 10'' by 10'' with a 2.5'' scale bar. North is up and East is left. Panels are color composites of $K$ (red), $y$ (green), $i$ (blue). Targets with broad H$\alpha$ are bordered in gold, other lines in green, and tentative H$\alpha$ lines in dashed blue. Object details are listed in Table \ref{Table: Targets}.}
    \label{Fig: Mosiac}
\end{figure*}

Spectroscopic follow-up observations were performed on November 2--3 (UT), 2025, using the Keck Near Infrared Echellette Spectrometer (NIRES) instrument \citep{wilson_mass_2004} with a slit width of 0.55$''$. We observed 8 targets with NIRES (PI: Labb\'e). Figure \ref{Fig: Mosiac} displays a mosaic view of the observed objects in $i$, $y$, and $K$. North is up and East is left in all images. Target RA, DEC, and photo-z are listed in Table \ref{Table: Targets} ($T_{obs}$ does not include bad frames).

The standard NIRES setup was used following \cite{wilson_mass_2004} to observe in the $y$, $J$, $H$, and $K$ bands simultaneously (across all orders). Flats and standard arcs (but no darks) were performed prior to the start of observations. Because the targets were so faint, blind-offset from a nearby star was required. In order to precisely observe the targets, offsets were calculated from stars identified in the original HSC catalogs used to identify the LRD candidates to ensure accurate relative astrometry. Specifically, as there was no robust star flag, we selected offset stars within 1' of each target by requiring $18<y<20.5$ and a point-like profile measured as the ratio of 1.0'' and 0.5'' apertures. 

The wavelength coverage of NIRES spans 0.94 to 2.45 $\mu$m, but with a detector gap from $1.855$--$1.888\ \mu$m. The seeing conditions ranged from 0.6--1.3", averaging 1" on the first night and 0.8" on the second. The phase of the Moon was Waxing Gibbous, approximately three days before full moon, reaching $\approx 70 ^{\circ}$ in elevation during the night and setting between 3:30 and 4:30am local time. Only objects in the DEEP2-F3 field were observed at $<30^\circ$ from the Moon. Each night consisted of approximately 10 hours of observing time.

Observations of telluric stars were taken to correct for telluric absorption from the atmosphere and to later provide flux calibration. Telluric observations were airmass matched to science targets and taken before and after science observations. A0V stars were observed with an ABBA dither pattern with an offset of 2.5'', 8 second exposures, 3 coadds, and 1 read per coadd. Observations of LRD candidates utilize an ABC dither pattern with a primary offset of 4'' (although 5'' was also used), 300 seconds of integration, 1 coadd, and 2 reads per coadd.

\begin{sidewaystable*}
    \centering
    \caption{Observed target details}
    \setlength{\tabcolsep}{4pt}
    \footnotesize
    \begin{tabular}{ccccccccccc}
        \toprule
        \textbf{Target}             & \textbf{Telescope} & \textbf{RA} & \textbf{Dec} & $z_{phot}$ & $z_{spec}$ & $K_{mag}$ & $T_{obs}\; [s]$ & Date & Lines & H$\alpha$ Coverage\\ 
        \midrule
        XMMLSS-AGN-2               & Keck+VLT & 02:26:05.816 & $-$04:29:03.077 & 2.18 & 2.135 & 20.02 & 4,800+7,300 & 11/02,12/13,14,17 & H$\alpha$, Ly$\alpha$ & yes \\
        XMMLSS-z2LRD1               & Keck     & 02:17:12.223 & $-$04:41:41.336 & 2.34 & 2.257 & 21.98 & 5,400       & 11/02             & H$\alpha$, [OIII] & yes \\
        XMMLSS-z2LRD2               & Keck     & 02:17:39.552 & $-$05:08:14.460 & 2.32 & 2.101 & 22.47 & 9,900       & 11/02,03          & H$\alpha$, [OIII] & yes \\
        XMMLSS-AGN-1                & VLT      & 02:19:11.608 & $-$05:15:09.237 & 3.41 & 2.437 & 21.17 & 7,300       & 12/14,15          & H$\alpha$ & yes \\
        ECOSMOS-z2LRD1              & Keck     & 09:59:26.488 & +01:38:46.015 & 2.38 & 2.391 & 22.47 & 9,900       & 11/02, 11/03      & H$\alpha$, [OIII] & yes \\
        DEEP23-z2LRD1               & Magellan & 23:27:35.881 & $-$00:13:41.654 & 2.29 & 2.261 & 22.09 & 9,638       & 10/18,19          & H$\alpha$, [OIII] & yes \\
        XMMLSS-LAE-2                & Keck+VLT & 02:20:19.285 & $-$03:49:11.173 & 2.18 & 2.350 & 22.04 & 2,400+2,920 & 11/02,12/16       & Ly$\alpha$ & yes \\
        XMMLSS-LAE-3                & VLT      & 02:24:24.023 & $-$04:19:29.428 & 2.51 & 2.940 & 22.46 & 2,920       & 12/15             & Ly$\alpha$ & no \\
        XMMLSS-LAE-4                & VLT      & 02:26:44.204 & $-$05:05:33.837 & 3.16 & 3.182 & 22.19 & 8,760       & 12/17             & Ly$\alpha$ & no \\
        XMMLSS-LAE-1                & VLT      & 02:17:27.071 & $-$05:17:29.330 & 2.70 & 2.598 & 22.04 & 2,920       & 12/18             & Ly$\alpha$ & yes \\
        DEEP23-NOLINE-2             & Keck     & 23:30:02.175 & $-$00:15:26.987 & 2.42 & --    & 22.67 & 2,400       & 11/03             & Tentative & uncertain \\
        XMMLSS-NOLINE-8             & Keck+VLT & 02:27:18.492 & $-$04:24:29.148 & 2.18 & --    & 22.37 & 3,600+2,920 & 11/03,12/15       & Tentative & uncertain \\
        XMMLSS-NOLINE-1             & VLT      & 02:18:36.791 & $-$04:09:50.222 & 2.73 & --    & 22.10 & 2,920       & 12/18             & - & uncertain \\
        XMMLSS-NOLINE-5             & VLT      & 02:24:13.688 & $-$04:40:10.871 & 3.02 & --    & 21.46 & 2,920       & 12/14             & - & uncertain \\
        ECOSMOS-NOLINE-3            & VLT      & 09:59:02.520 & +00:56:54.938 & 2.53 & --    & 22.31 & 2,920       & 12/15             & - & uncertain \\
        ECOSMOS-NOLINE-1            & VLT      & 09:55:31.989 & +03:20:11.998 & 2.83 & --    & 22.74 & 745         & 12/15             & - & uncertain \\
        ECOSMOS-NOLINE-4            & VLT      & 10:04:37.544 & +03:24:02.455 & 2.58 & --    & 21.80 & 4,380       & 12/15             & - & uncertain \\
        XMMLSS-NOLINE-6             & VLT      & 02:26:24.854 & $-$05:02:50.966 & 2.62 & --    & 22.20 & 2,920       & 12/15,16          & - & uncertain \\
        XMMLSS-NOLINE-2             & VLT      & 02:19:11.927 & $-$05:26:24.871 & 2.92 & --    & 22.17 & 2,920       & 12/16             & - & uncertain \\
        XMMLSS-NOLINE-7             & VLT      & 02:26:40.103 & $-$04:33:26.584 & 2.38 & --    & 22.30 & 2,920       & 12/16             & - & uncertain \\
        XMMLSS-NOLINE-3             & VLT      & 02:21:48.858 & $-$04:36:22.026 & 2.65 & --    & 22.37 & 2,920       & 12/16             & - & uncertain \\
        XMMLSS-NOLINE-4             & VLT      & 02:22:42.106 & $-$04:40:44.892 & 3.00 & --    & 22.43 & 2,920       & 12/17             & - & uncertain \\
        ECOSMOS-NOLINE-2            & VLT      & 09:57:50.695 & +01:42:45.813 & 2.11 & --    & 22.44 & 2,920       & 12/17             & - & uncertain \\
        DEEP23-NOLINE-1             & Keck     & 23:26:29.722 & $-$00:24:26.411 & 2.24 & --    & 21.99 & 4,500       & 11/03             & - & uncertain \\
        \botrule
    \end{tabular}
    \tablecomments{$K_{mag}$ is AB, in VISTA $K_s$ or UKIRT $K$ depending on the field. $T_{obs}$ is on-source time. Exposure times denote separate Keck and VLT exposures, not a total. H$\alpha$ Coverage is yes or no where $z_{spec}$ places H$\alpha$ inside or outside the spectral coverage, and uncertain where only $z_{phot}$ is available. Tentative lines are too low SNR to confirm; see Figure \ref{Appendix: Appendix Spectra}}
    \label{Table: Targets}
\end{sidewaystable*}

\subsection{Keck Reduction}

The spectra were reduced using \texttt{\href{https://pypeit.readthedocs.io/en/stable/}{PypeIt}} \citep{prochaska_pypeit_2020} version 1.18.1, which is a Python package and pipeline designed for spectroscopic data reduction. \texttt{PypeIt} supports many telescopes and instrument modes, including Keck/NIRES. \texttt{PypeIt} performs end-to-end processing of raw frames to science ready outputs. 

Before processing with \texttt{PypeIt} was performed, all raw frames were inspected to approximate the spatial and spectral location of emission features. This was done manually by image differencing coadded frames of the same type (e.g. A$-$B). A \texttt{PypeIt} configuration file was then generated to define the set of science exposures, telluric standard observations, and flat-field frames to be reduced. Science frames were set to \texttt{arc} and \texttt{tilt} to use OH skylines for wavelength calibration. Because the targets of observation were very faint, manual extraction was performed instead of using \texttt{PypeIt}'s automatic object finding algorithm. B-spline fitting was also turned off.

All the individual 2D differenced science pairs (e.g. A--B) were then coadded together using inverse variance weighting. Weights were calculated by inverse variance weighting of each default 1D spectrum output by \texttt{PypeIt} around the H$\alpha$ profile and then normalized to unity. None of \texttt{PypeIt}'s weighting options could optimally coadd the 2D pairs, only a given list of weights did. The individual pairs were then coadded together with another manual trace and FWHM with which \texttt{PypeIt} extracted the coadded 1D spectrum. A telluric standard star was then used to derive a sensitivity function which was applied to flux the coadded 1D extraction. Because the targets were acquired by blind offset, slit losses can differ between the science target and the standard, so each coadded spectrum was scaled with \texttt{PypeIt} to its catalog $K$ magnitude (Table \ref{Table: Targets}).

\subsection{VLT/X-shooter} \label{Sec: Data XShooter}

Targets from our sample were observed with X-shooter on the VLT \citep{vernet_xshooter_2011} under programme 116.294D (PI: Matthee), carried out on 2025 December 13--18. The instrument configuration, observing strategy, and reduction are described in more detail by \cite{torralba_black_2026} and we refer the reader to that paper for the full observing setup.

X-shooter records the UVB, VIS and NIR arms simultaneously. Slit widths of $1.0"$, $0.9"$ and $0.9"$ were used, giving nominal resolving powers of $R = 5400$, $8900$ and $5600$ respectively. Targets were acquired by blind offset from a nearby reference star, as they are too faint for direct acquisition. Observations used a standard ABBA nod pattern along the slit with $4\arcsec$ throws and $1\arcsec$ jitters in the NIR arm to improve sky subtraction.

The data were reduced with the ESO X-shooter pipeline and telluric-corrected as part of the program's reduction. We take the resulting order-merged, flux-calibrated and telluric-corrected 1D spectra, one per observing block and arm.

\subsection{Magellan/FIRE} \label{Sec: Data FIRE}

DEEP23-z2LRD1 was observed with FIRE on the 6.5m Magellan Baade telescope on 2024 October 18--19 for 9.6 ks, with a slit width of 0.75$''$, the sample-up-the-ramp readout mode and an ABBA nod pattern. The observations and data reduction are further described by \citet{ma_counting_2025}. The data were reduced with \texttt{PypeIt} \citep{prochaska_pypeit_2020}.

\subsection{JWST/NIRSpec} \label{Sec: Data NIRSpec}

XMMLSS-z2LRD2 was observed with JWST/NIRSpec in program GO-3567 (PI: Valentino; \citealp{ito_deepdive_2026}). The spectrum is publicly available on the DAWN JWST Archive (DJA), processed with \texttt{msaexp} \citep{brammer_msaexp_2023, degraaff_rubies_2025, heintz_strong_2024}. The observation uses the G235M/F170LP grating with a total exposure time of $\approx 10,500$ s. The extracted spectrum covers $1.61-4.23\ \mu$m with a gap at $1.90-2.07\ \mu$m. At $z = 2.1005$ that gap contains H$\alpha$ ($2.035\ \mu$m), which our NIRES observation captures, while Pa$\beta$ falls at an observed $3.974\ \mu$m. XMMLSS-z2LRD2 was also imaged by \cite{dunlop_primer_2021} in JWST programme GO-1837 (PRIMER, PI: Dunlop) and by \cite{muzzin_minerva_2025} in GO-7814 (MINERVA, PI: Muzzin). The top panel of Figure \ref{Fig: NIRSpec XMM4} shows the full rest-frame spectrum, and the bottom-left panel an RGB image stamp (reduced with \texttt{grizli}: \citealp{brammer_grizli_2023}) with the NIRSpec MSA slitlet overlaid.

\subsection{Chandra Observations} \label{Sec: Chandra Data}

We searched the Chandra Data Archive to find potential X-ray emission of our candidates. We searched at the RA and DEC of each target in Table \ref{Table: Targets} and retrieved all primary and secondary packages within a 10 arcmin radius. Seven targets have usable observations: XMMLSS-AGN-1 (PI: Pacucci 24700302), XMMLSS-LAE-1, XMMLSS-z2LRD2 (PI: Hasinger 16900937), XMMLSS-NOLINE-5 (PI: Maughan 17800360), XMMLSS-LAE-3 (PI: Jones 06800443), ECOSMOS-z2LRD1 and ECOSMOS-NOLINE-2 (PI: Civano 14900904). Observations of XMMLSS-NOLINE-4, XMMLSS-NOLINE-7, XMMLSS-NOLINE-8, and DEEP23-z2LRD1 are within 10' but not on any chip. Only XMMLSS-NOLINE-5 is observed close to on-axis with ACIS-S, all other observations are off-axis, five with ACIS-I and one more with ACIS-S.

We processed the Chandra observations with \texttt{CIAO v4.18} \citep{fruscione_ciao_2006}, reprocessing each ObsID with \texttt{chandra\_repro} and \texttt{CALDB} 4.12.3 to apply the latest calibration. To correct the astrometry, we detected sources in each ObsID with \texttt{fluximage} and \texttt{wavdetect} and matched them to Gaia DR3 sources within $12'$ of the aimpoint. Using \texttt{wcs\_match}, we matched pairs within $2''$ while rejecting pairs with post-fit residuals $>1''$. We applied each shift to both the aspect solution and the event file with \texttt{wcs\_update}. We then combined the corrected event files of each target with \texttt{merge\_obs}, to make the counts images in Figure \ref{Fig: X-ray} and to detect sources (Section \ref{Sec: X-ray}).

\subsection{Comparison Samples} \label{Sec: Data Comparison}

Reddened quasars are the population most likely to contaminate a ground-based LRD selection at these redshifts. We compare our candidates to typical red-quasars from the literature using the F2M sample \citep{glikman_first2mass_2012}, heavily reddened quasars (HRQs; \citealt{stepney_rest_2026}), as well as JWST identified LRDs \citep{degraaff_little_2025}. We select objects from these samples at $2<z<3$. The F2M sample \citep{glikman_first2mass_2004} was specifically designed to target highly obscured AGN by requiring cross matched radio from FIRST \citep{becker_first_1995}, infrared from 2MASS, and faint optical observations \citep{skrutskie_two_2006}. Of the 120 F2M red quasars of \cite{glikman_first2mass_2012}, 15 are in the redshift range $2 <z < 3$, 9 of which have spectra covering H$\alpha$. We use the available photometry for HRQs from \cite{stepney_rest_2026} only as the spectra are not publicly available: 49 are in the redshift range $2 < z < 3$, with E(B$-$V) $=0.4-1.8$. From \cite{degraaff_little_2025}, there are 8 LRDs in the same redshift range.

\section{Results} \label{Sec: Results}

\begin{figure*}
    \centering
    \includegraphics[width=\linewidth]{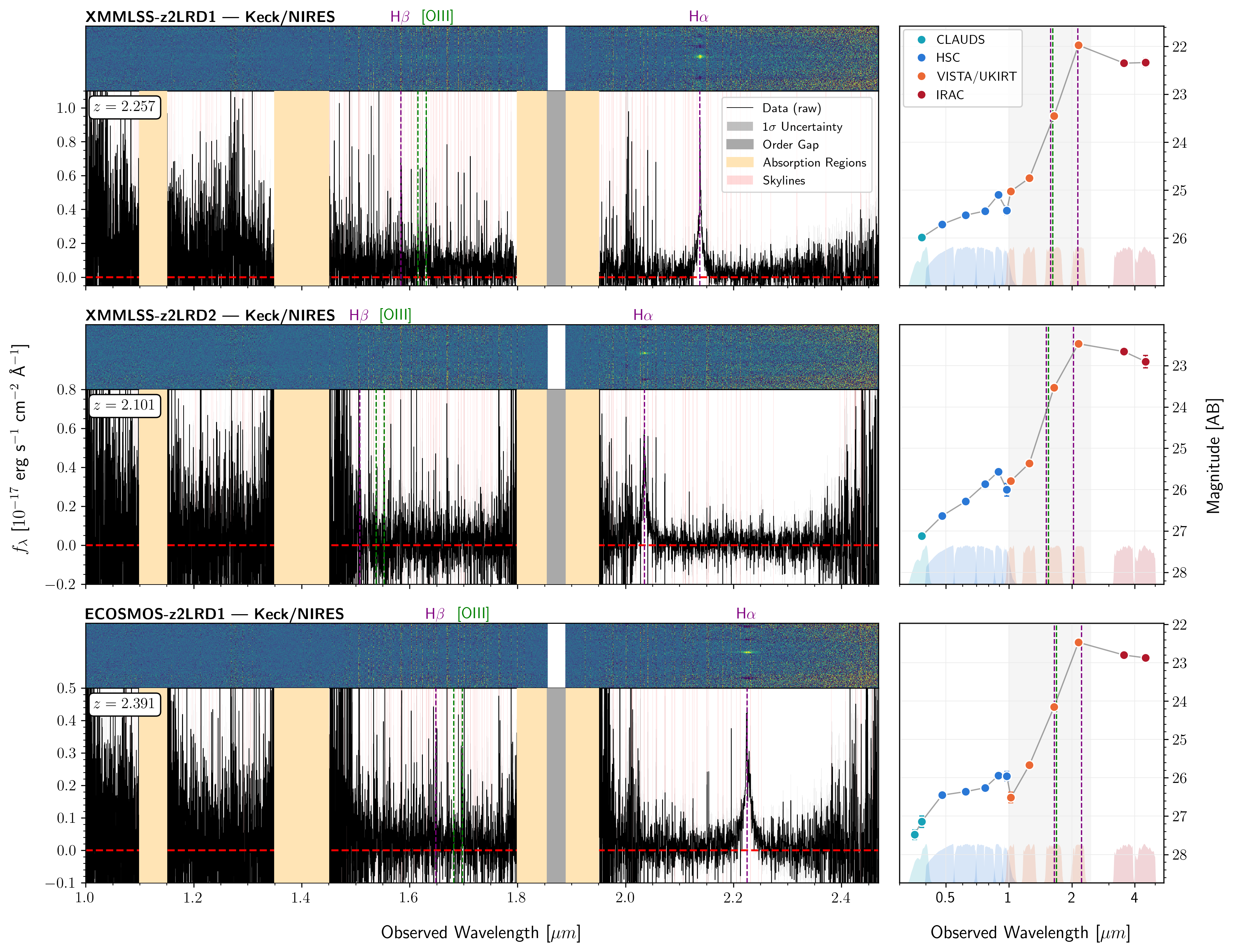}
    \caption{Keck/NIRES Spectroscopy of XMMLSS-z2LRD1, XMMLSS-z2LRD2, and ECOSMOS-z2LRD1. Top panels show the stitched 2D Spectrum with the 1D in the bottom panel. Target spectra are shown in black with uncertainty in light gray. Prominent emission lines are labeled with a vertical dashed line representing the center. A gap in the detector exists from $1.855 -1.888\ \mu$m seen in dark gray. Absorption bands are colored orange. Skylines visible in the 2D are marked in the 1D as vertical red bars. The right column shows the observed-frame photometric SED of each target in AB magnitudes, colored by survey (CLAUDS $u$, HSC $grizy$, VISTA/UKIRT $YJHK$ and Spitzer/IRAC $3.6$, $4.5\ \mu$m) with the filter transmission curves along the bottom. Vertical dashed lines mark the location of H$\alpha$ and H$\beta$ (purple), and [OIII] (green). The gray band marks the wavelength range of the spectrum.}
    \label{Fig: Spectra NIRES}
\end{figure*}

\begin{figure*}
    \centering
    \includegraphics[width=\linewidth]{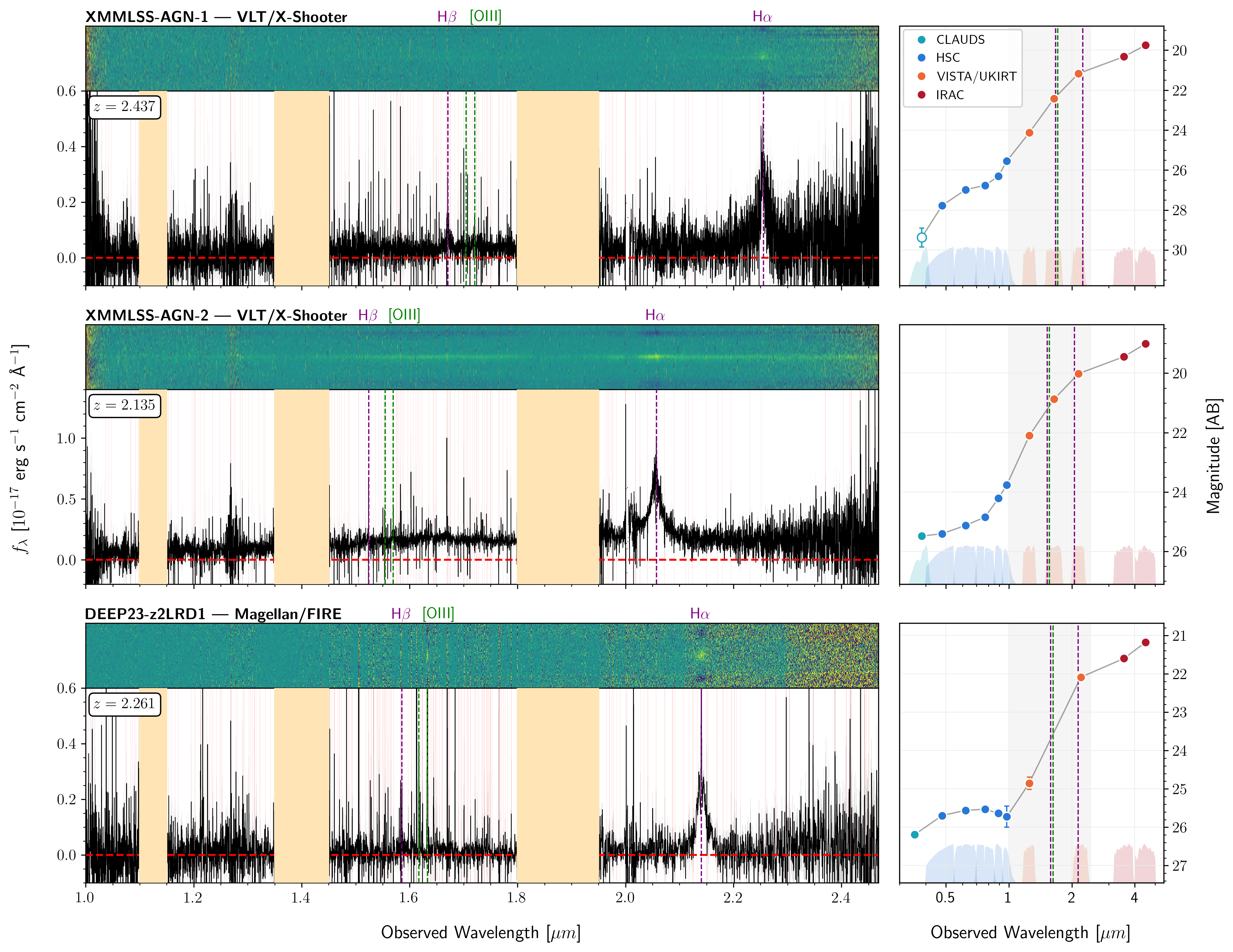}
    \caption{As in Figure \ref{Fig: Spectra NIRES}. Spectroscopy of the candidates XMMLSS-AGN-1, XMMLSS-AGN-2, and DEEP23-z2LRD1.}
    \label{Fig: Spectra VLT}
\end{figure*}

Figures \ref{Fig: Spectra NIRES} and \ref{Fig: Spectra VLT} display the results of the reductions explained in Section \ref{Sec: Data}. The top panel of each sub-figure shows the rectified 2D spectrum. Persistence from telluric star observations is identifiable by three horizontal lines coinciding with the ABBA telluric dither pattern. Three features are detected in the slits: H$\alpha$, Ly$\alpha$, and [OIII].

The bottom panel of each sub-figure shows the coadded 1D spectrum. The $1 \sigma$ uncertainty is shown in light gray. Identified lines are labeled and include a colored vertical dashed line. Areas of strong atmospheric absorption are recorded from 1.10--1.15, 1.35--1.45, and 1.80--1.95 $\mu$m corresponding with a drop in detector throughput and telluric transmission. These are shown as the light orange filled regions. A gap in the detector occurs from 1.855-1.888\,$\mu$m shown as the dark gray filled region. Strong OH emission from the sky is highlighted in red where pixels are flagged as having $>50\%$ higher noise than the local median ($3 \times$ the spectrum median shown).

Of the 24 observed targets, 10 have spectroscopic redshifts, but only 6 show detected H$\alpha$. These have a median on-source time of $\approx 8,500$ s against $2,920$ s for the targets without H$\alpha$. This comparison is biased, however: targets were only observed for longer than $\approx$ 1 hour if H$\alpha$ was detected. To test whether depth alone explains the difference, we take XMMLSS-z2LRD2, which has the weakest broad H$\alpha$ of the six detections, and degrade its spectrum to a $2,920$ s exposure by inflating the error array by $\sqrt{9900/2920}$ and adding the corresponding noise. Across 500 noise realizations the broad line is recovered in every one, so depth alone does not account for the difference. The implied detection limit of a broad line is $L_{\mathrm{H}\alpha} \approx 3 \times 10^{42}$ erg s$^{-1}$ at $z = 2.1$ ($\approx 6 \times 10^{42}$ erg s$^{-1}$ at $z = 2.9$). Spectroscopic redshifts place H$\alpha$ beyond the coverage for XMMLSS-LAE-3 and XMMLSS-LAE-4 (Appendix \ref{Sec: Appendix}). For the 14 targets with only photometric redshifts, coverage is uncertain (Table \ref{Table: Targets}); taken at face value, four would fall beyond it. For the rest, there are two possibilities: H$\alpha$ lies within the coverage and must be intrinsically fainter than the limit above, or non-detections are due to photometric redshift errors.

\begin{figure*}
    \centering
    \includegraphics[width=0.99\linewidth]{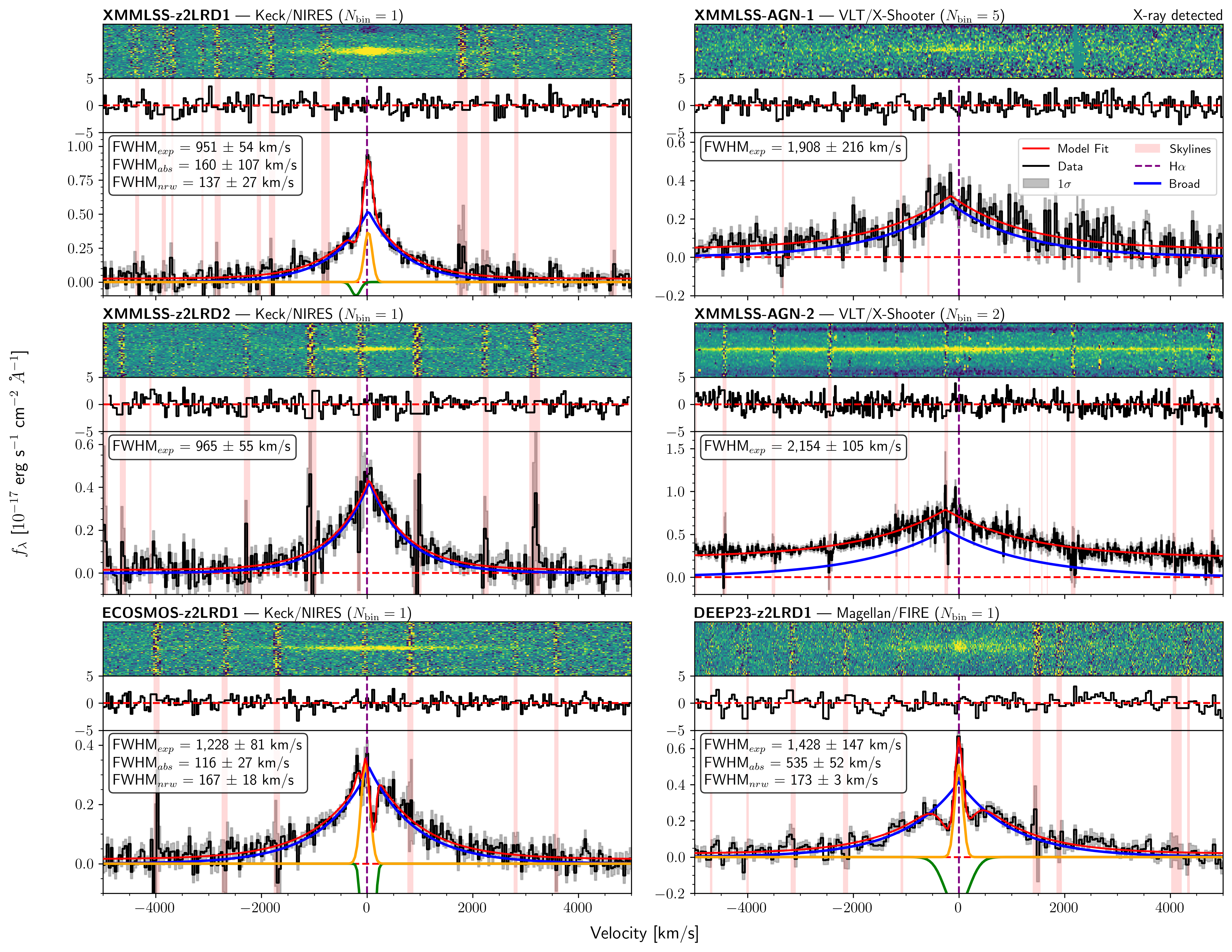}
    \caption{H$\alpha$ emission line fits for each candidate. The broad component is exponential for all targets (Table \ref{Table: Components}). For each source, the top panel shows the 2D spectrum, the middle panel shows the normalised residuals, and the lower panel shows the spectrum (black) with $1\sigma$ uncertainties (grey). The total model fit is shown in red, with individual components shown separately: the broad component in blue, the narrow component in orange, and the absorption component in green. Pixels coinciding with skyline regions (red shading) are excluded from the fit. The FWHM of each fitted component is listed in the upper left of each panel.}
    \label{Fig: Velocity}
\end{figure*}
\subsection{Line Modelling} \label{Sec: Line Modelling}

Since $M_{\rm BH}$, $L_{\rm bol}$, and $\lambda_{\rm Edd}$ are all derived from the H$\alpha$ line profile, accurate modeling and robust uncertainty estimates are critical to assessing whether these sources are consistent with supermassive black hole accretion.  We utilise the \texttt{curve\_fit} routine from \texttt{SciPy} \citep{virtanen_scipy_2020} to fit the models, performing weighted least squares fitting which optimizes model parameters by minimizing the residuals weighted by the per-pixel flux uncertainties. Uncertainties on the fitted parameters are returned by the fit \citep[evaluated with \texttt{NumPy}:][]{harris_array_2020}, which we then propagate into derived quantities.

Radiative transfer effects in dense gas can significantly alter hydrogen-line profiles and the physical properties inferred from them \citep{chang_impact_2026}. We therefore tested different line shapes (exponential, Gaussian) and combinations of broad, narrow, absorption, and continuum components \citep{madau_little_2026,rusakov_little_2026}. Each fit includes a free constant continuum term optimized simultaneously with the line components. Model performance was evaluated using the Bayesian Information Criterion (BIC), computed for each fit and used to identify the preferred model. Differences are interpreted as $\Delta \mathrm{BIC} > 10$ is strong evidence to reject a model. All emission components share the same systemic velocity, which sets $z_{\rm spec}$ (Table \ref{Table: Targets}). They are tied because leaving each emission component free allows it to compensate for the profile shape elsewhere in the model, mostly against the absorption component. The absorber itself keeps a free centroid to allow for the possibility of inflows or outflows. ECOSMOS-z2LRD1 is the one exception as with tied centroids the narrow component disappears entirely. 

We tie the narrow H$\alpha$ width to that of [OIII] for all objects that have it, using the [OIII] widths measured in Figure \ref{Fig: Oiii Hb fits}. Narrow H$\alpha$ is assumed to be embedded in a broad line whereas [OIII] measures the same width with no degeneracy. When left free, the narrow H$\alpha$ width is consistent with [OIII]. We do not adopt this constraint for XMMLSS-z2LRD1, whose [OIII] is $73 \pm 33$ km s$^{-1}$ broader than narrow H$\alpha$ if their widths are untied. Fixing narrow H$\alpha$ to [OIII] forces the absorption component to absorb more flux and shift towards systemic to compensate.

The H$\alpha$ emission line fits are presented in Figure \ref{Fig: Velocity}, with the FWHM of each component listed in the upper left of each panel. The fitted parameters of every component are in Table \ref{Table: Components}, and the $\Delta \mathrm{BIC}$ of every model tested is in Table \ref{Table: BIC}. Broad components are colored blue, narrow in orange, and absorption in green. All six fits share the constraints described above: the emission components share a single centroid (except in ECOSMOS-z2LRD1), the narrow width is tied to [OIII], and a free continuum term is fitted alongside them. The objects are described individually below. 

\begin{itemize}
    \item XMMLSS-AGN-2 is best fit by a single exponential ($\mathrm{FWHM}=2\ln2\tau$) profile. Adding an absorption component is not significantly preferred ($\Delta \mathrm{BIC}=3.6$), and adding a narrow component is ruled out ($\Delta \mathrm{BIC}>10$). The $180$--$320$ km/s range, where a bad pixel overlays the object trace, is excluded from the fit. XMMLSS-AGN-2 does show clear double-peaked Ly$\alpha$ (Figure \ref{Fig: Lyman Lines}).

    \item XMMLSS-z2LRD1 is best described by an exponential+narrow+absorption model, but the absorber is tentative. The BIC does not distinguish it from the exponential+narrow model ($\Delta \mathrm{BIC} < 2$). 
    
    \item XMMLSS-z2LRD2 is best fit by a single exponential component. Adding a narrow component is slightly disfavored ($\Delta \mathrm{BIC}=5.2$) and is not ruled out (Table \ref{Table: BIC}). 
    
    \item XMMLSS-AGN-1 is best fit by a single exponential. Adding a narrow component ($\Delta \mathrm{BIC}=3.2$) or an absorber ($\Delta \mathrm{BIC}=7.6$) is not significantly preferred, and a Gaussian broad+narrow model is not ruled out ($\Delta \mathrm{BIC}=8.4$). It also has the largest offset between photometric and spectroscopic redshift of the ten targets with spectroscopic redshifts ($z_{\rm phot}=3.41$, $z_{\rm spec}=2.437$; Table \ref{Table: Targets}).
    
    \item ECOSMOS-z2LRD1 is formally best fit by an exponential+absorption model, but a clear narrow component is visible, so we adopt the exponential+narrow+absorption model ($\Delta \mathrm{BIC}=2.6$). Its absorber is redshifted by $46 \pm 14$ km s$^{-1}$ relative to the systemic. An OH line falling on the narrow emission is very weak and left included in the fit. 
    
    \item DEEP23-z2LRD1 is best fit with an exponential+narrow+absorption model with no support for other component combinations.
\end{itemize}

\begin{table*}[ht!]
    \centering
    \caption{Components of the adopted H$\alpha$ model for each target, with [OIII]$\lambda$5007 for comparison. The components listed are those of the adopted model and the broad profile shape is given in parentheses. Velocities are relative to systemic, marked $\equiv 0$; narrow and broad $H\alpha$ are tied in case of degenerate fits; for ECOSMOS-z2LRD1, whose emission centroids are fitted independently, systemic is the narrow centroid. FWHM values are corrected for the instrumental resolution. $F$ is the observed flux of that component and $L$ the corresponding luminosity. $^{\dagger}$ marks a width held at the [OIII] value.}
    \begin{tabular}{llcccc}
        \toprule
        \textbf{Target} & \textbf{Component} & $\Delta v$ \textbf{[km/s]} & \textbf{FWHM [km/s]} & $F$ $10^{-16}$ \textbf{[erg s$^{-1}$ cm$^{-2}$]} & $L$ $10^{42}$ \textbf{[erg s$^{-1}$]} \\
        \midrule
        \textbf{XMMLSS-AGN-2} & Broad (exponential) & $\equiv 0$ & $2,154 \pm 105$ & $11.9 \pm 0.6$ & $40 \pm 2$ \\
        \midrule
        \textbf{XMMLSS-z2LRD1} & Broad (exponential) & $\equiv 0$ & $951 \pm 54$ & $5.2 \pm 0.2$ & $20.1 \pm 0.7$ \\
         & Narrow & $\equiv 0$ & $137 \pm 27$ & $0.49 \pm 0.07$ & $1.9 \pm 0.3$ \\
         & Absorption & $-233 \pm 35$ & $160 \pm 107$ & $-0.14 \pm 0.08$ & $-0.6 \pm 0.3$ \\
         & [OIII]$\lambda$5007 & $-10 \pm 13$ & $210 \pm 19$ & $1.09 \pm 0.08$ & $4.2 \pm 0.3$ \\
        \midrule
        \textbf{XMMLSS-z2LRD2} & Broad (exponential) & $\equiv 0$ & $965 \pm 55$ & $4.0 \pm 0.2$ & $13.1 \pm 0.6$ \\
         & [OIII]$\lambda$5007 & $-65 \pm 23$ & $119 \pm 41$ & $0.36 \pm 0.06$ & $1.2 \pm 0.2$ \\
        \midrule
        \textbf{XMMLSS-AGN-1} & Broad (exponential) & $\equiv 0$ & $1,908 \pm 216$ & $5.8 \pm 0.7$ & $27 \pm 3$ \\
        \midrule
        \textbf{ECOSMOS-z2LRD1} & Broad (exponential) & $70 \pm 22$ & $1,228 \pm 81$ & $4.3 \pm 0.2$ & $19.1 \pm 0.9$ \\
         & Narrow & $\equiv 0$ & $167 \pm 18$$^{\dagger}$ & $0.54 \pm 0.07$ & $2.4 \pm 0.3$ \\
         & Absorption & $46 \pm 14$ & $116 \pm 27$ & $-0.7 \pm 0.2$ & $-3.1 \pm 0.7$ \\
         & [OIII]$\lambda$5007 & $47 \pm 11$ & $167 \pm 18$ & $0.28 \pm 0.02$ & $1.24 \pm 0.09$ \\
        \midrule
        \textbf{DEEP23-z2LRD1} & Broad (exponential) & $\equiv 0$ & $1,428 \pm 147$ & $6.1 \pm 0.3$ & $24 \pm 1$ \\
         & Narrow & $\equiv 0$ & $173 \pm 3$$^{\dagger}$ & $0.70 \pm 0.05$ & $2.7 \pm 0.2$ \\
         & Absorption & $-37 \pm 12$ & $535 \pm 52$ & $-1.2 \pm 0.2$ & $-4.6 \pm 0.7$ \\
         & [OIII]$\lambda$5007 & $10 \pm 1$ & $173 \pm 3$ & $1.12 \pm 0.01$ & $4.37 \pm 0.04$ \\
        \botrule
    \end{tabular}
    \label{Table: Components}
\end{table*}

\begin{table*}[ht!]
    \centering
    \caption{$\Delta$BIC of each H$\alpha$ model with respect to the lowest-BIC model for that target. E is an exponential broad component, G a Gaussian broad component, N a narrow component and A an absorption component. Models with $\Delta BIC < 10$, i.e.\ not ruled out with very strong evidence, are in bold. $^{\dagger}$ marks the adopted model, discussed in the text.}
    \begin{tabular*}{\textwidth}{@{\extracolsep{\fill}}l rrrrrrrr}
        \toprule
        \textbf{Target} & $\Delta$BIC$_{\rm E}$ & $\Delta$BIC$_{\rm E+A}$ & $\Delta$BIC$_{\rm E+N}$ & $\Delta$BIC$_{\rm E+N+A}$ & $\Delta$BIC$_{\rm G}$ & $\Delta$BIC$_{\rm G+A}$ & $\Delta$BIC$_{\rm G+N}$ & $\Delta$BIC$_{\rm G+N+A}$ \\
        \midrule
        XMMLSS-AGN-2     & \textbf{0.0}\rlap{$^{\dagger}$} & \textbf{3.6} & 12.4 & 13.4 & 50.7 & 45.7 & 17.3 & 20.9 \\
        XMMLSS-z2LRD1     & 166.5 & 89.0 & \textbf{1.0} & \textbf{0.0}\rlap{$^{\dagger}$} & 551.0 & 423.9 & 38.8 & 30.0 \\
        XMMLSS-z2LRD2     & \textbf{0.0}\rlap{$^{\dagger}$} & 10.8 & \textbf{5.2} & 16.5 & 74.9 & 70.5 & 12.9 & 17.6 \\
        XMMLSS-AGN-1    & \textbf{0.0}\rlap{$^{\dagger}$} & \textbf{7.6} & \textbf{3.2} & 11.6 & 29.4 & 32.2 & \textbf{8.4} & 15.9 \\
        ECOSMOS-z2LRD1  & 62.6 & \textbf{0.0} & 60.1 & \textbf{2.6}\rlap{$^{\dagger}$} & 88.2 & 67.6 & 25.9 & 27.2 \\
        DEEP23-z2LRD1 & 356.4 & 48.5 & 58.7 & \textbf{0.0}\rlap{$^{\dagger}$} & 491.4 & 408.6 & 48.4 & 29.9 \\
        \botrule
    \end{tabular*}
    \label{Table: BIC}
\end{table*}

Figure \ref{Fig: Oiii Hb fits} shows the results of H$\beta$ and [OIII] fitting process. XMMLSS-z2LRD1 [OIII] FWHM is larger than narrow H$\alpha$: excess blueshifted, broad $\lambda$5007 flux is observed. Modeling this additional component shows positive but not strong evidence against a broad [OIII] outflow component ($\Delta \mathrm{BIC}=4.9$). Higher resolution spectroscopy is required to confirm this component.

An OH line falls on the [OIII] profile of XMMLSS-z2LRD1, although its impact is negligible because it does not mask any important fitting features. Notably, H$\beta$ is not detected in any of the six sources, resulting in Balmer decrement ratios that deviate strongly from the Case B recombination value \citep{osterbrock_astrophysics_2006}. [OIII] emission is also not detected in XMMLSS-AGN-1 or XMMLSS-AGN-2. This may be due to the low signal-to-noise of these two sources.

The Balmer Decrement and [OIII]/H$\beta$ ratio will be quantified further and discussed in Section \ref{Sec: Ionisation}. Tentative H$\alpha$ lines are given for targets DEEP23-NOLINE-2 and XMMLSS-NOLINE-8 in Appendix~\ref{Sec: Appendix}.

The broad-line FWHM and flux (Table \ref{Table: Components}) are used to derive $M_{\rm BH}$, $L_{\rm bol}$, and $\lambda_{\rm Edd}$ in Section \ref{Sec: BHM}. 
\subsection{Continuum Slopes}
\label{Sec: Beta Slopes}

\begin{figure*}[t!]
    \centering
    \includegraphics[width=0.49\linewidth]{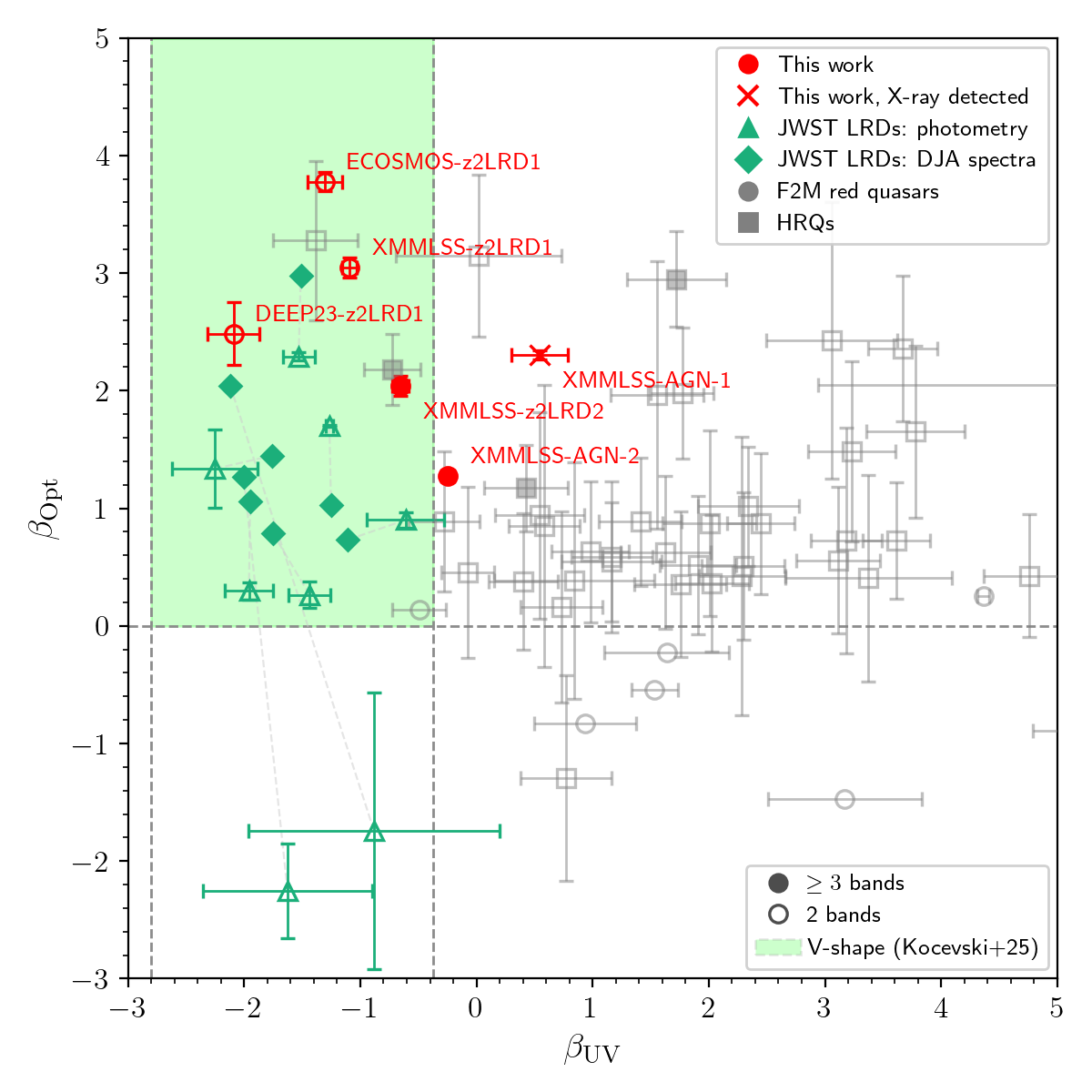}
    \hfill
    \includegraphics[width=0.49\linewidth]{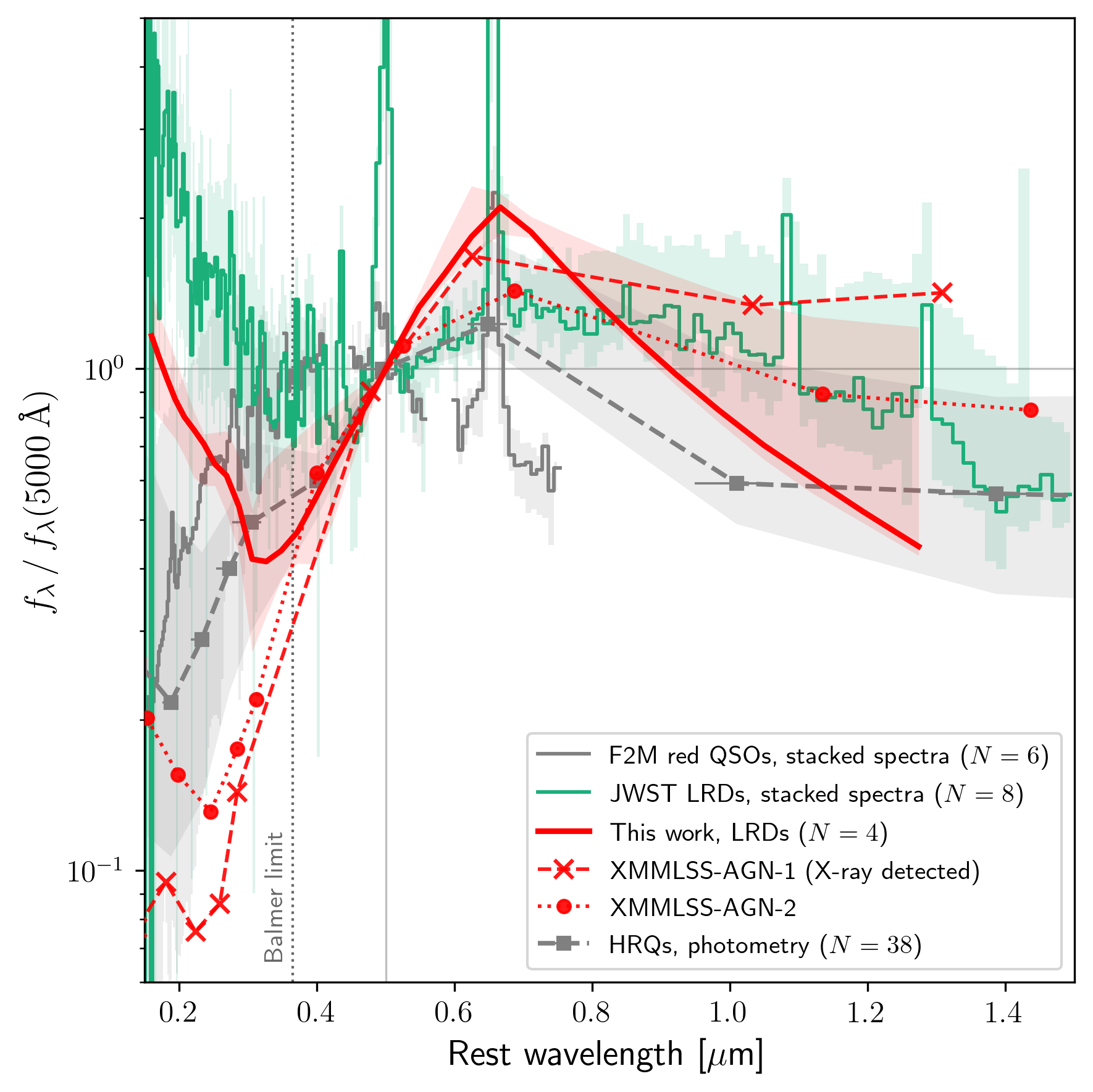}
    \caption{Two comparisons of the LRD candidates against reddened quasars. \textit{Left:} continuum slope fits $\beta_{\rm UV}$ and $\beta_{\rm opt}$ comparing LRD candidates in this work, JWST LRDs \citep{degraaff_little_2025}, typical red-quasars \citep{glikman_first2mass_2012}, and heavily reddened quasars \citep{stepney_rest_2026}. Objects with 3 or more bands on both sides are marked colour filled, other objects are marked with open symbols. The green filled box marks the selection used by \cite{kocevski_rise_2025} to identify LRDs. \textit{Right:} rest-frame SEDs normalised to $f_\lambda$ at $5000 \textrm{\AA}$ and spectra normalised over $5100$--$5300 \textrm{\AA}$. Our four LRDs are shown in red as the median of their photometric SEDs, shaded over their full range. The X-ray detected XMMLSS-AGN-1 (crosses, dashed) and XMMLSS-AGN-2 (circles, dotted) are shown separately. JWST LRDs (green) and F2M red quasars (grey, solid) are median spectra stacks of the objects in the left panel. The HRQs (grey, dashed) have photometry only and are shown as a band median.}
    \label{Fig: Beta Slopes}
\end{figure*}

V-shape is a core component of identifying LRDs, consistently observed close to the Balmer limit ($\lambda_{\rm rest} \approx 3645\ \textrm{\AA}$) \citep{setton_little_2024, rinaldi_not_2025}. We follow the criteria set out by \cite{kocevski_rise_2025} in identifying LRDs based on their V-shape:

\begin{equation*}
    \begin{split}
            -2.8 <\ & \beta_{\rm UV} < -0.37\ \mathrm{AND}\ \\ 
            & \beta_{\rm opt} > 0\ \mathrm{AND} \\
            & \mathrm{SNR}_{\mathrm{band}} \geq 3
    \end{split}
\end{equation*}

We calculate $\beta_{\rm UV}$ and $\beta_{\rm opt}$ based on photometry by fitting $\log_{10}(f_{\lambda}) = \beta \log_{10}(\lambda)+c$ using the same routine described in Section \ref{Sec: Line Modelling} and fitted independently over the rest-frame windows $1500 < \lambda_{\mathrm{rest}} < 3646\ \textrm{\AA}$ and $3646 < \lambda_{\mathrm{rest}} < 7500\ \textrm{\AA}$ for $\beta_{\mathrm{UV}}$ and $\beta_{\mathrm{Opt}}$ respectively. 

In every case the slopes are fitted to broadband photometry, never to the spectra directly, but the source of that photometry differs between samples (described in Section \ref{Sec: Data Comparison}). All objects are placed on a common HSC $griz$ plus VISTA $YJHK_{\rm s}$ system where possible. For our Keck/NIRES, VLT/X-shooter and Magellan/FIRE candidates the band fluxes are the catalogue photometry (Section \ref{Sec: Data Photometric}). For the F2M quasars and the JWST LRDs, which have spectra but no photometry on this system, the band fluxes are synthesised from the spectra. For the HRQs we fit the published photometry of \cite{stepney_rest_2026} directly.

HSC and VISTA bandpass photometry is synthesised from the F2M and JWST spectra with \texttt{sedpy} \citep{johnson_bdj_2021} through the same filter curves, so that every object is measured through the same bandpasses. Photometric bands covering the Balmer limit are excluded from all samples. For the synthesised photometry we additionally require that at least 90 per cent of a filter's throughput lies over sampled spectral data, which rejects bands truncated by the end of a spectrum or falling in a gap. A slope is measured only where at least two bands survive in a window. Since our observed targets use real catalogue photometry, we do not mask emission lines in the spectra which we derive our synthetic photometry from in order to keep the comparisons consistent. Uncertainties on the synthesised band fluxes are obtained from 200 Monte Carlo realisations in which the spectrum is perturbed by its per-pixel errors and the photometry re-synthesised. The F2M spectra carry no per-pixel error array, so band uncertainties for this sample cannot be measured. The HRQs of \cite{stepney_rest_2026} have mixed photometric fluxes from SDSS, DES, Pan-STARRS, HSC, and KiDS, so this sample does not use only HSC filters as the others do.

The results of the slope fitting are presented in the left panel of Figure \ref{Fig: Beta Slopes}. The same objects appear as rest-frame SEDs in the right panel, normalised at $5000 \textrm{\AA}$. Objects fit with three or more bands on both sides are colour filled, others are marked as open symbols. The 3 Keck/NIRES candidates and DEEP23-z2LRD1 fall in the green region used by \cite{kocevski_rise_2025} to select LRDs. The two VLT/X-shooter candidates do not meet the slope criteria, although XMMLSS-AGN-2 is only just outside the UV slope limit, while XMMLSS-AGN-1 lies well outside it. XMMLSS-AGN-2 also has a strong \textit{Spitzer}/MIPS 24\,$\mu$m detection. Our colour selection (Section \ref{Sec: Data Selection}) is looser than the \cite{kocevski_rise_2025} criteria in the UV, so meeting them is not guaranteed by selection. Few of the reddened quasars meet the criteria: one of the seven F2M red quasars with usable photometry and two of the 38 HRQs of \cite{stepney_rest_2026}. The JWST LRDs (green triangles) are measured from photometry synthesised from their spectra; the green diamonds are the same objects' slopes as published by the DJA from fits to the spectra themselves, joined to their triangles by faint dashed lines.
If lines are masked before synthesising the photometry, the scatter of $\beta_{\rm opt}$ shows improved agreement for five of eight JWST objects. For most objects only two bands fall redward of the Balmer limit, so $\beta_{\rm opt}$ is determined entirely by the flux ratio between those two bands, with no degrees of freedom. We attribute the scatter between our synthetic photometry and the DJA spectroscopic slopes to this sampling.
\subsection{Ionisation and Balmer Decrements} \label{Sec: Ionisation}

\begin{figure*}
    \centering
    \includegraphics[width=\linewidth]{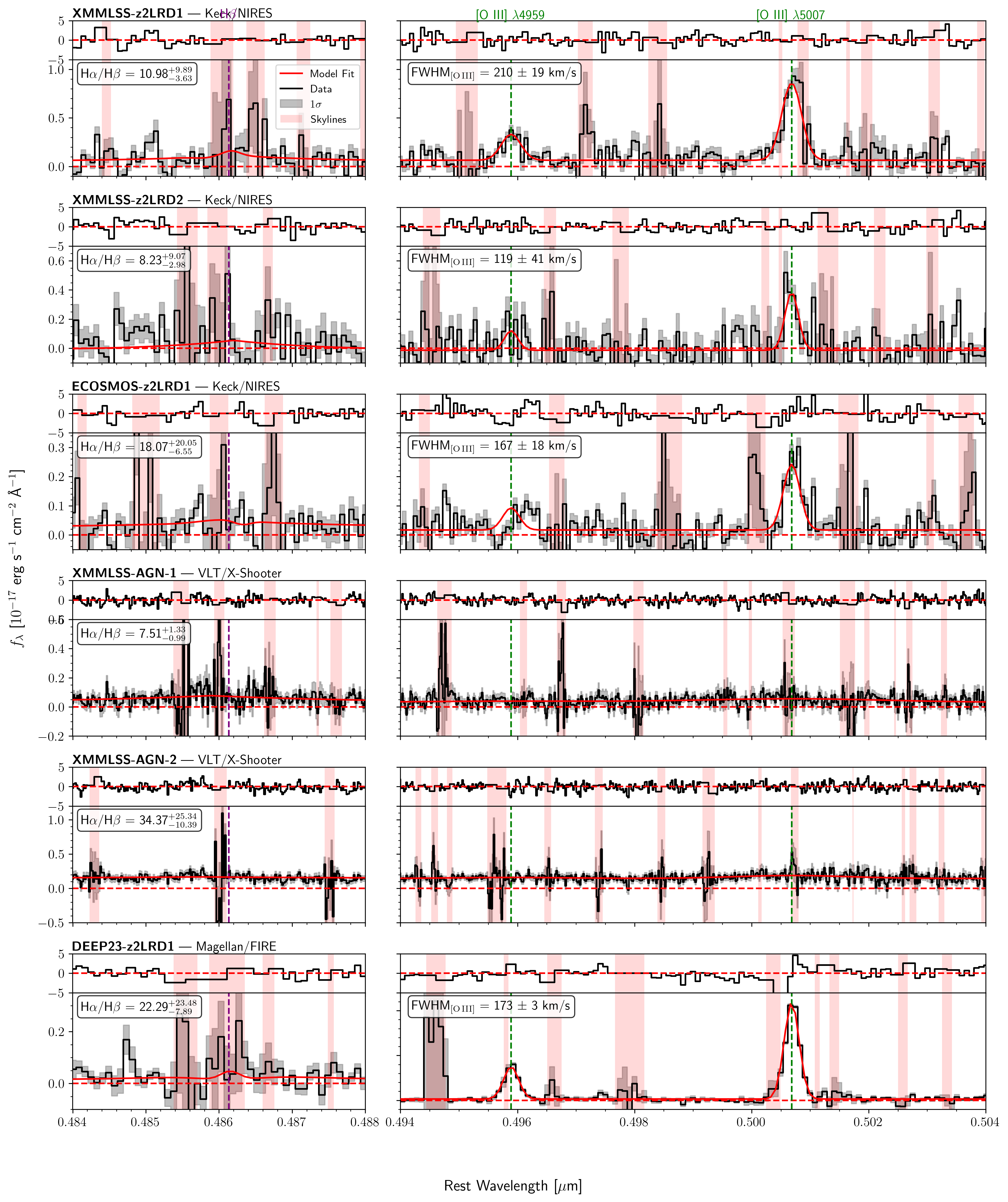}
        \caption{Fits to the H$\beta$ (left) and [OIII] $\lambda\lambda4959,5007$ (right) emission line regions for the six sources with broad H$\alpha$. In each panel, the black line shows the observed spectrum with the grey shaded region indicating the $1\sigma$ error, the red curve shows the best-fit model (for H$\beta$, the adopted H$\alpha$ model scaled by a single normalisation; for [OIII], Gaussians), and red shaded bands mark regions of skyline contamination. Residuals are shown in the upper sub-panels. Dashed vertical lines indicate the expected positions of H$\beta$ (purple) and [OIII] (green) of each source. The measured FWHM of the [OIII] fit is annotated in each right panel. H$\beta$ is not detected in any source. Each left panel is annotated with H$\alpha$/H$\beta$, as listed in Table \ref{Table: Line Ratios}.}
    \label{Fig: Oiii Hb fits}
\end{figure*}

The only diagnostic available to us is the [OIII]$\lambda$5007/H$\beta$ ratio. However, as shown in Figure \ref{Fig: Oiii Hb fits}, H$\beta$ is undetected in all 6 candidates. To estimate the H$\beta$ flux, we fix the parameters to the values determined from the H$\alpha$ fit (translated to the H$\beta$ rest wavelength) and fit a single overall flux normalisation. We assume the same relative contribution of each component.

\begin{table}
    \centering
    \caption{Measured emission-line ratios for the sample. Values are the median of Monte Carlo draws of the fitted amplitudes, with uncertainties from the 16th and 84th percentiles. $\Delta v$ is the velocity of [OIII]$\lambda5007$ relative to narrow H$\alpha$.}
    \label{Table: Line Ratios}
    \setlength{\tabcolsep}{4pt}
    \begin{tabular}{lccc}
        \hline
        Object & [OIII]/H$\beta$ & H$\alpha$/H$\beta$ & $\Delta v$ \\
         & & & [km s$^{-1}$] \\
        \hline
        XMMLSS-AGN-2    & -- & $34.37^{+25.34}_{-10.39}$ & -- \\
        XMMLSS-z2LRD1    & $2.16^{+1.97}_{-0.73}$ & $10.98^{+9.89}_{-3.63}$ & $-10 \pm 13$ \\
        XMMLSS-z2LRD2    & $0.72^{+0.81}_{-0.28}$ & $8.23^{+9.07}_{-2.98}$ & -- \\
        XMMLSS-AGN-1   & -- & $7.51^{+1.33}_{-0.99}$ & -- \\
        ECOSMOS-z2LRD1 & $1.10^{+1.23}_{-0.40}$ & $18.07^{+20.05}_{-6.55}$ & $+47 \pm 11$ \\
        DEEP23-z2LRD1 & $4.36^{+4.61}_{-1.54}$ & $22.29^{+23.48}_{-7.89}$ & $+10 \pm 1$ \\
        \hline
    \end{tabular}
\end{table}

Table \ref{Table: Line Ratios} shows the emission line ratios of our sample. We measure total Balmer decrements well above the standard Case B recombination value of H$\alpha$/H$\beta \approx 2.86$ \citep{osterbrock_astrophysics_2006}. Every source lies far above the Case B value, so the elevation itself is secure. 
\subsection{X-ray Overlap} \label{Sec: X-ray}

\begin{figure}[ht!]
    \centering
    \includegraphics[width=\columnwidth]{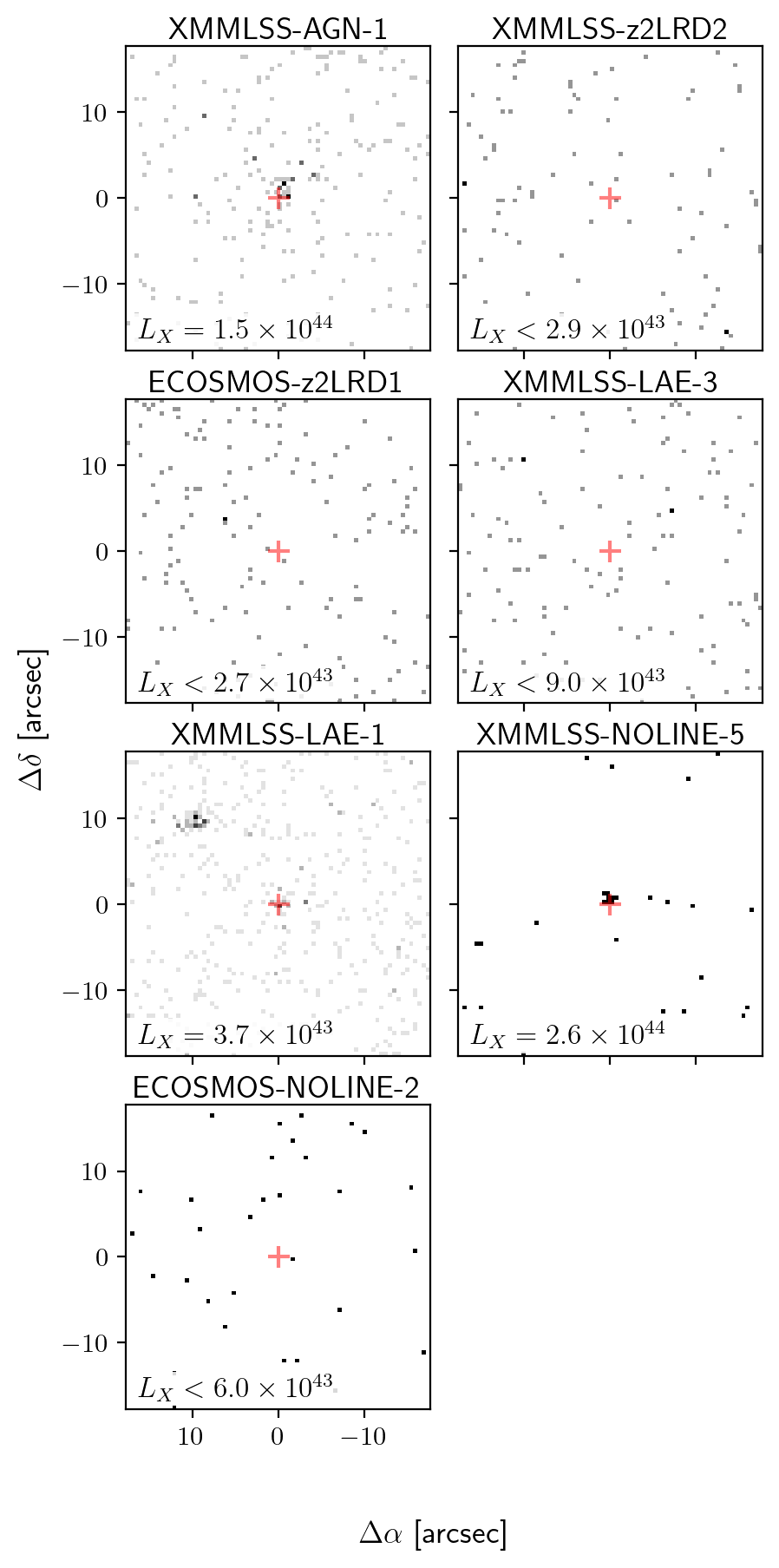}
    \caption{Merged Chandra $0.5-7$ keV counts images of the seven targets with archival coverage, built from the corrected event files. Each panel is $36'' \times 36''$ and centred on the optical position (red cross), with North up and East left. The rest-frame $2-10$ keV luminosity, or $3\sigma$ upper limit for undetected targets, is given in erg s$^{-1}$ in each panel.}
    \label{Fig: X-ray}
\end{figure}

Chandra observations of our targets were introduced with a description of the reduction in Section \ref{Sec: Chandra Data}. We measured X-ray photometry at the optical position of each target and ObsId with \texttt{srcflux} in the soft ($0.5-1.2$ keV), medium ($1.2-2$ keV), hard ($2-7$ keV) and broad ($0.5-7$ keV) bands with a circular aperture enclosing 90\% of the PSF at 1 keV. \texttt{srcflux} then combines the counts and exposures of all ObsIDs covering a target into a single count rate and flux. Fluxes assume a power law with $\Gamma=1.8$, no intrinsic absorption, and Galactic $N_H$ retrieved by \texttt{colden}. We convert the observed $0.5-7$ keV flux to rest-frame $2-10$ keV luminosity as $L_X=4\pi d_L^2F_{0.5-7\ \mathrm{keV}} K$, where $K$ is the k-correction for the same power law from \texttt{calc\_kcorr}. The $1\sigma$ uncertainties and $3\sigma$ upper limits come from two runs of \texttt{srcflux}, at 68.27\% and 99.87\% confidence respectively. We detect sources with \texttt{wavdetect} on the merged image of each target and consider a source detected if it lies within the 90\% PSF radius of its optical position. For detected targets we give fluxes with $1\sigma$ uncertainties, and for undetected targets $3\sigma$ upper limits. Results are given in Table \ref{Table: X-ray}, and the X-ray images at the position of each target are shown in Figure \ref{Fig: X-ray}. 

\begin{table*}[ht!]
    \centering
    \caption{Chandra X-ray fluxes and luminosities. Bold rows give the result from all ObsIDs combined. Fluxes are unabsorbed, in units of $10^{-15}$ erg s$^{-1}$ cm$^{-2}$, in the soft ($0.5-1.2$ keV), medium ($1.2-2$ keV) and hard ($2-7$ keV) bands. $L_X$ is the rest-frame $2-10$ keV luminosity from the $0.5-7$ keV flux. For targets detected by \texttt{wavdetect}, the combined values are given with $1\sigma$ uncertainties. All other values are $3\sigma$ upper limits. --: no measurement, because the target falls off the detector or there are no source or background counts.}
    \label{Table: X-ray}
    \begin{tabular}{lcccccc}
        \hline
        ObsID & Exp. (ks) & $\theta$ (arcmin) & $F_{\rm soft}$ & $F_{\rm medium}$ & $F_{\rm hard}$ & $L_X$ ($10^{42}$ erg s$^{-1}$) \\
        \hline
        \textbf{XMMLSS-AGN-1} & 227.9 &  & $2.12^{+1.65}_{-1.28}$ & $0.77^{+0.30}_{-0.25}$ & $3.42^{+0.90}_{-0.80}$ & $149^{+31}_{-28}$ \\
        \quad 26790 & 11.9 & 7.7 & $<52.51$ & $<7.53$ & $<37.43$ & $<953$ \\
        \quad 27162 & 12.9 & 7.9 & $<79.39$ & $<8.35$ & $<22.28$ & $<620$ \\
        \quad 27163 & 13.5 & 7.8 & $<64.28$ & $<7.60$ & $<18.13$ & $<495$ \\
        \quad 27164 & 14.9 & 7.8 & $<79.70$ & $<7.05$ & $<33.55$ & $<1010$ \\
        \quad 27165 & 11.2 & 7.8 & $<70.32$ & $<11.53$ & $<16.93$ & $<565$ \\
        \quad 27166 & 14.9 & 7.8 & $<52.83$ & $<10.44$ & $<19.28$ & $<686$ \\
        \quad 27167 & 13.5 & 7.8 & $<67.73$ & $<12.71$ & $<34.96$ & $<1151$ \\
        \quad 27168 & 10.2 & 7.8 & $<126.07$ & $<15.65$ & $<33.97$ & $<1423$ \\
        \quad 28985 & 20.8 & 7.8 & $<79.15$ & $<9.30$ & $<18.53$ & $<873$ \\
        \quad 28986 & 14.9 & 7.8 & $<58.91$ & $<9.30$ & $<16.67$ & $<537$ \\
        \quad 29004 & 10.9 & 7.8 & $<71.64$ & $<9.04$ & $<22.08$ & $<598$ \\
        \quad 29005 & 9.5 & 7.8 & $<85.24$ & $<13.87$ & $<26.22$ & $<854$ \\
        \quad 29021 & 14.2 & 7.8 & $<42.62$ & $<7.83$ & $<29.93$ & $<835$ \\
        \quad 29022 & 14.2 & 7.8 & $<63.54$ & $<9.99$ & $<17.95$ & $<591$ \\
        \quad 29023 & 13.5 & 7.8 & $<67.50$ & $<8.05$ & $<22.42$ & $<600$ \\
        \quad 29024 & 13.1 & 7.8 & $<73.04$ & $<8.70$ & $<20.30$ & $<555$ \\
        \quad 29025 & 13.6 & 7.8 & $<88.66$ & $<12.62$ & $<22.34$ & $<916$ \\
        \hline
        \textbf{XMMLSS-z2LRD2} & 100.4 &  & $<1.30$ & $<0.54$ & $<2.39$ & $<29$ \\
        \quad 17299 & 49.3 & 2.2 & -- & $<1.29$ & $<5.64$ & $<71$ \\
        \quad 17309 & 51.1 & 3.2 & $<2.26$ & $<0.94$ & $<3.09$ & $<39$ \\
        \hline
        \textbf{ECOSMOS-z2LRD1} & 198.7 &  & $<0.99$ & $<0.30$ & $<1.39$ & $<27$ \\
        \quad 15222 & 49.4 & 7.1 & $<2.51$ & $<1.15$ & $<6.37$ & $<98$ \\
        \quad 15223 & 50.9 & 9.1 & $<3.52$ & $<1.04$ & $<4.42$ & $<81$ \\
        \quad 15246 & 48.9 & 3.9 & $<1.82$ & $<1.20$ & $<3.20$ & $<63$ \\
        \quad 15247 & 49.6 & 6.3 & $<3.45$ & $<1.02$ & $<4.28$ & $<88$ \\
        \hline
        \textbf{XMMLSS-LAE-3} & 115.9 &  & $<0.70$ & $<0.39$ & $<4.61$ & $<90$ \\
        \quad 6390 & 11.9 & 8.5 & \multicolumn{4}{c}{--} \\
        \quad 6394 & 17.5 & 7.9 & $<5.09$ & $<2.60$ & $<15.82$ & $<367$ \\
        \quad 7182 & 22.9 & 7.9 & $<3.85$ & $<1.99$ & $<10.59$ & $<241$ \\
        \quad 7183 & 19.9 & 7.9 & $<4.39$ & $<2.28$ & $<14.08$ & $<323$ \\
        \quad 7184 & 22.7 & 7.9 & $<3.89$ & $<2.01$ & $<14.22$ & $<316$ \\
        \quad 7185 & 32.9 & 6.7 & $<2.06$ & $<1.39$ & $<7.71$ & $<155$ \\
        \hline
        \textbf{XMMLSS-LAE-1} & 385.6 &  & $<0.62$ & $0.14^{+0.09}_{-0.07}$ & $1.38^{+0.46}_{-0.42}$ & $37^{+12}_{-11}$ \\
        \quad 17287 & 47.5 & 9.3 & \multicolumn{4}{c}{--} \\
        \quad 17288 & 48.6 & 4.1 & $<3.39$ & $<1.39$ & $<9.70$ & $<194$ \\
        \quad 17289 & 46.0 & 2.6 & $<3.26$ & $<1.61$ & $<6.26$ & $<179$ \\
        \quad 17290 & 47.5 & 8.4 & $<7.37$ & $<2.08$ & $<9.19$ & $<250$ \\
        \quad 17292 & 49.1 & 9.4 & \multicolumn{4}{c}{--} \\
        \quad 17293 & 49.4 & 4.3 & $<2.72$ & $<1.48$ & $<5.77$ & $<135$ \\
        \quad 17294 & 49.4 & 3.0 & $<2.34$ & $<1.54$ & $<5.00$ & $<133$ \\
        \quad 17295 & 46.0 & 8.4 & $<7.15$ & $<1.88$ & $<10.06$ & $<250$ \\
        \quad 17298 & 49.3 & 8.3 & $<4.78$ & $<2.27$ & $<7.76$ & $<221$ \\
        \quad 17299 & 49.3 & 7.6 & $<5.26$ & $<1.51$ & $<7.28$ & $<189$ \\
        \hline
        \textbf{XMMLSS-NOLINE-5} & 11.9 &  & $1.79^{+1.71}_{-1.04}$ & $1.89^{+1.15}_{-0.81}$ & $<12.03$ & $265^{+129}_{-98}$ \\
        \quad 18262 & 11.9 & 1.3 & $1.79^{+1.71}_{-1.04}$ & $1.89^{+1.15}_{-0.81}$ & $<12.03$ & $265^{+129}_{-98}$ \\
        \hline
        \textbf{ECOSMOS-NOLINE-2} & 47.5 &  & $<2.52$ & $<1.55$ & $<4.09$ & $<60$ \\
        \quad 15244 & 47.5 & 5.2 & $<2.52$ & $<1.55$ & $<4.09$ & $<60$ \\
        \hline
    \end{tabular}
\end{table*}

Seven targets fall on the detector in archival Chandra observations (Section \ref{Sec: Chandra Data}). XMMLSS-AGN-1, which has broad H$\alpha$, is detected, giving $L_X = (1.5\pm0.3)\times10^{44}$ erg s$^{-1}$. XMMLSS-LAE-1 is also detected, with $L_X = (3.7^{+1.2}_{-1.1})\times10^{43}$ erg s$^{-1}$. XMMLSS-NOLINE-5 is detected, giving $L_X = (2.6^{+1.3}_{-1.0})\times10^{44}$ erg s$^{-1}$. XMMLSS-LAE-3 is undetected, with a $3\sigma$ upper limit of $L_X < 9.0\times10^{43}$ erg s$^{-1}$. ECOSMOS-NOLINE-2 is also undetected, with $L_X < 6.0\times10^{43}$ erg s$^{-1}$.

XMMLSS-z2LRD2 and ECOSMOS-z2LRD1 also have broad H$\alpha$ and are both undetected, with $3\sigma$ upper limits of $L_X < 2.9\times10^{43}$ and $< 2.7\times10^{43}$ erg s$^{-1}$ respectively. XMMLSS-z2LRD2 has only a single count within its extraction aperture. An inverse-variance weighted stack of the two gives $L_X < 2.0\times10^{43}$ erg s$^{-1}$. Assessing whether these luminosities are weak requires a comparison against the broad H$\alpha$ luminosity, which we defer to Section \ref{Sec: Cosmic Noon LRDs}.
\subsection{NIRSpec Pa$\beta$ of XMMLSS-z2LRD2} \label{Sec: NIRSpec}

\begin{figure*}
    \centering
    \includegraphics[width=\textwidth]{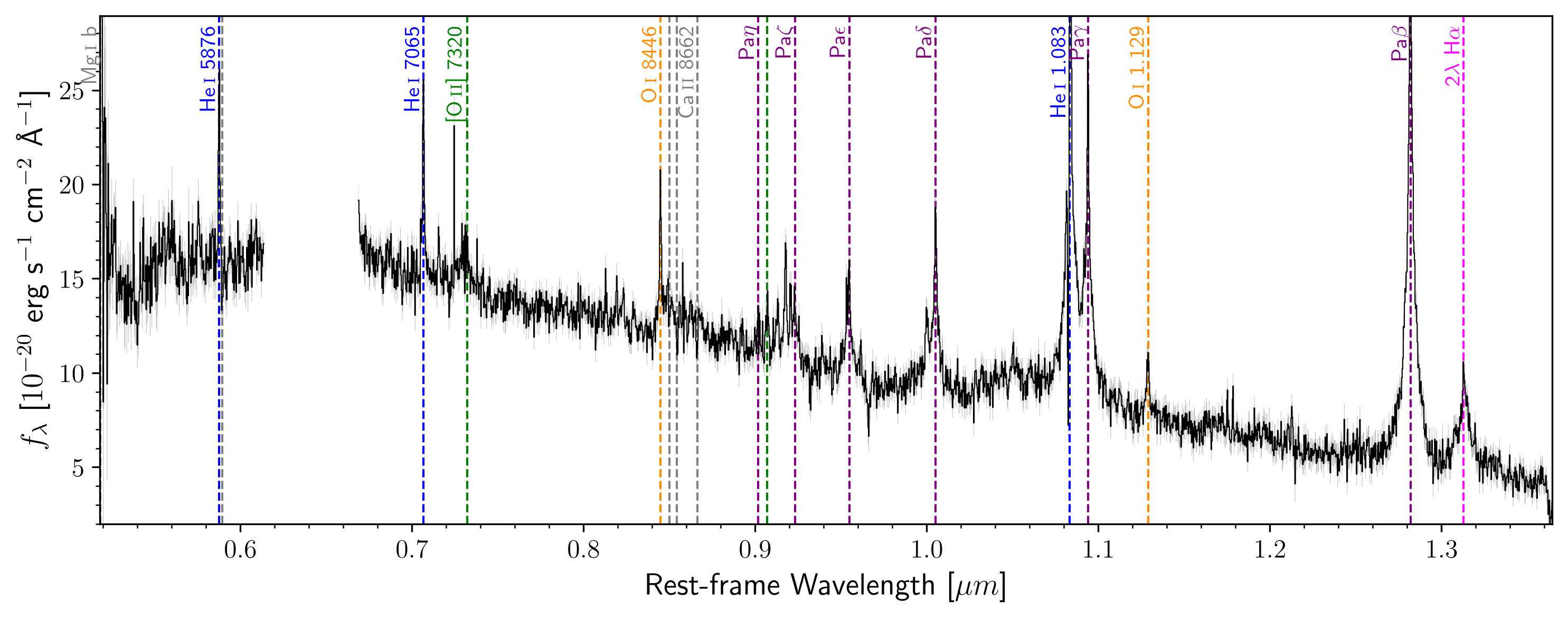} \\
    \vspace{0.3em}
    \begin{minipage}{0.32\textwidth}
        \centering
        \includegraphics[width=\linewidth]{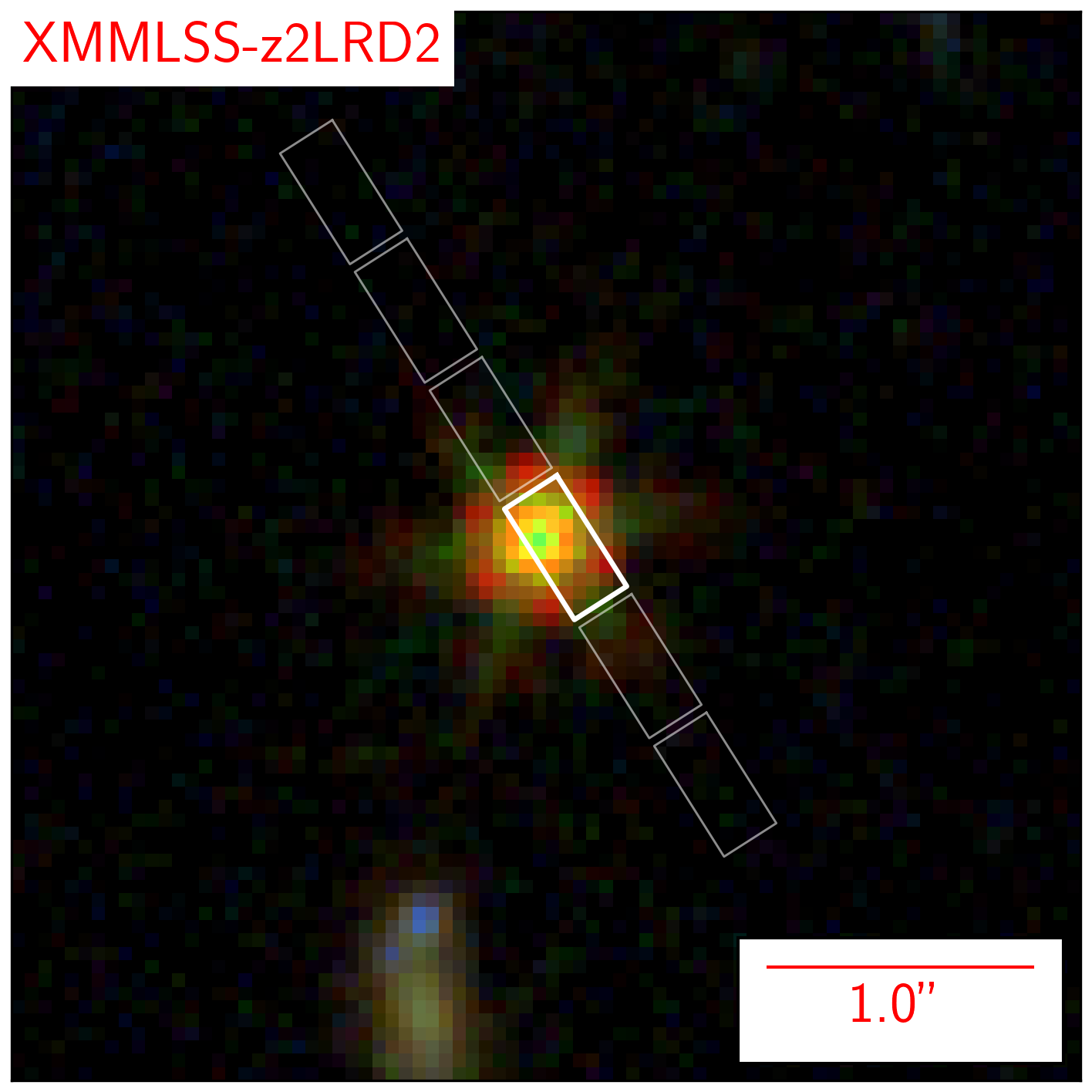}
    \end{minipage}
    \hfill
    \begin{minipage}{0.32\textwidth}
        \centering
        \includegraphics[width=\linewidth]{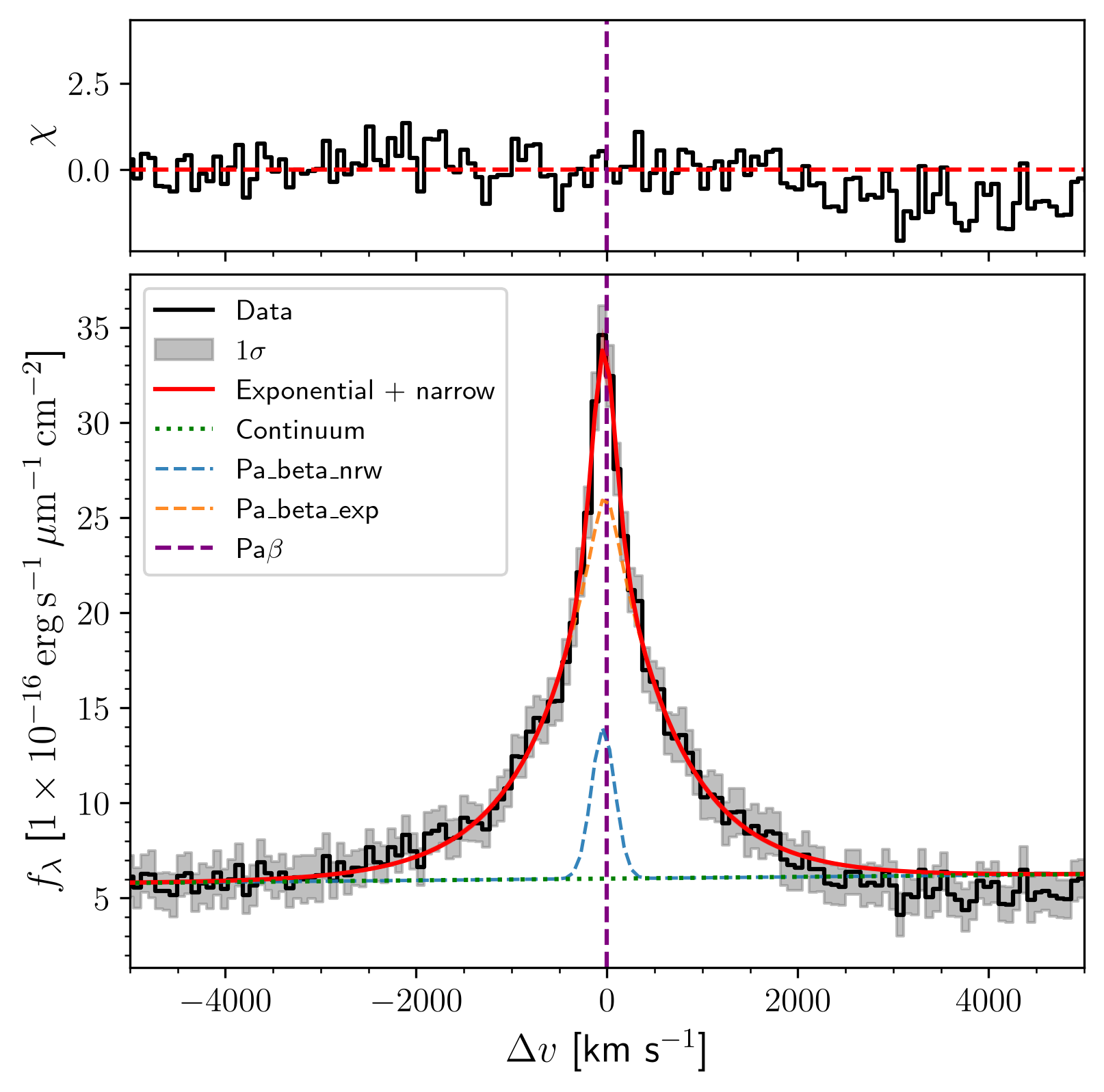}
    \end{minipage}
    \hfill
    \begin{minipage}{0.32\textwidth}
        \centering
        \includegraphics[width=\linewidth]{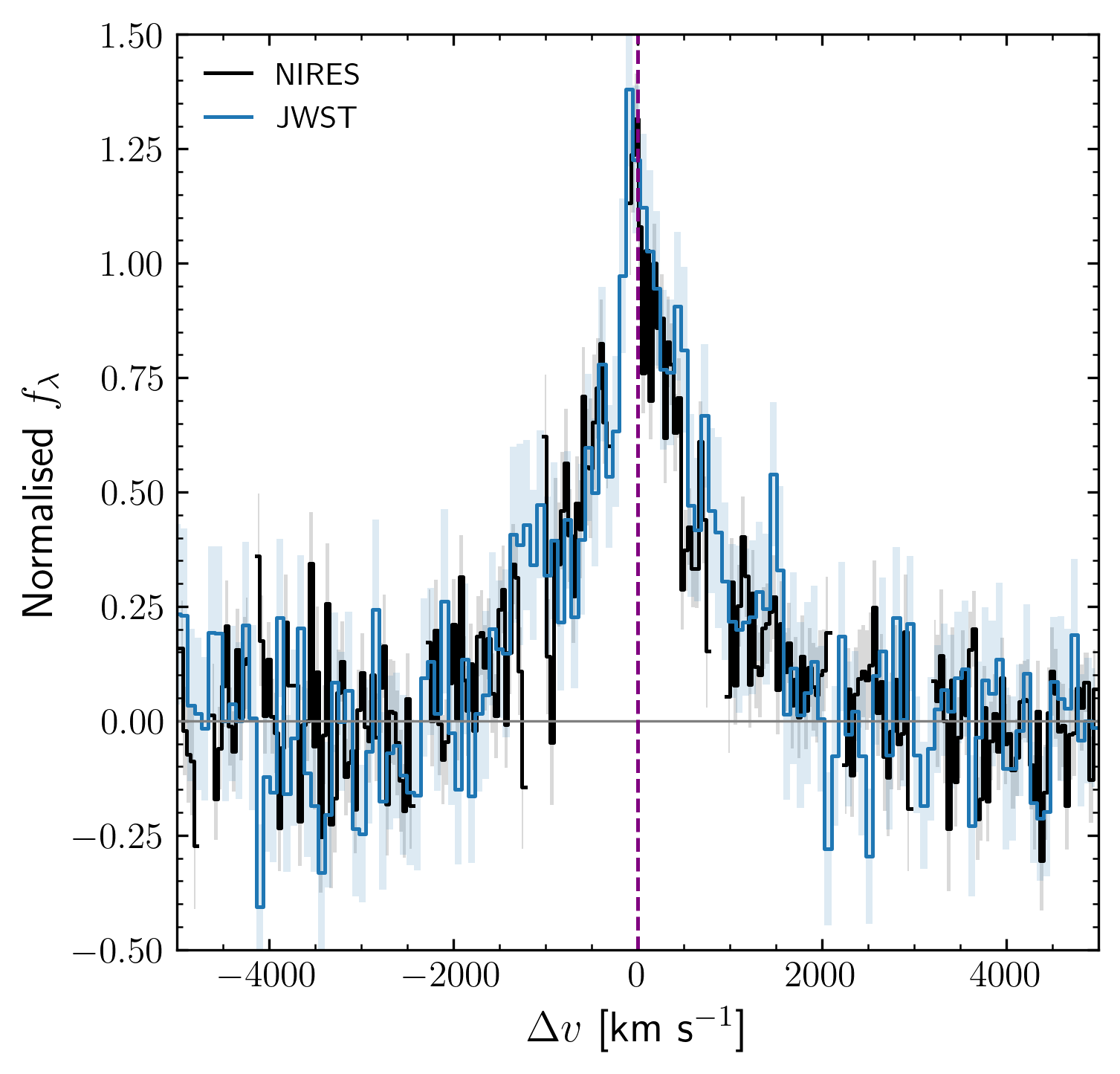}
    \end{minipage}
    \caption{\textit{Top:} the rest-frame NIRSpec G235M/F170LP spectrum of XMMLSS-z2LRD2. The gap at $0.61-0.67\ \mu$m is the NRS1/NRS2 detector gap, which contains H$\alpha$. However, H$\alpha$ is also seen in second order, marked 2$\lambda$ H$\alpha$. \textit{Bottom left:} NIRCam RGB of XMMLSS-z2LRD2 using PRIMER imaging. F090W+F115W+F150W in blue, F200W+F277W in green, and F356W+F410M+F444W in red. The NIRSpec MSA slitlet is outlined in white, with the shutter containing the source drawn solid and the background shutters faint. \textit{Bottom middle:} the adopted Pa$\beta$ fit, an exponential broad component plus a narrow Gaussian over a local linear continuum, fitted with \texttt{unite}. The spectrum is in black with its $1\sigma$ uncertainty in grey, the total model in red, the continuum dotted green, and the two components dashed. Normalised residuals are shown above. Velocities are relative to Pa$\beta$ at the fitted redshift. \textit{Bottom right:} the Keck/NIRES H$\alpha$ (black) overlaid on the second-order (2$\lambda$) H$\alpha$ from JWST/NIRSpec (blue). Each spectrum is normalised by its median flux.}
    \label{Fig: NIRSpec XMM4}
\end{figure*}

Pa$\beta$ is detected in the NIRSpec spectrum of XMMLSS-z2LRD2 (Section \ref{Sec: Data NIRSpec}), giving a second broad line to compare against the NIRES H$\alpha$. The bottom-middle panel of Figure \ref{Fig: NIRSpec XMM4} shows the fit.

We fit Pa$\beta$ with \textbf{U}nified li\textbf{N}e \textbf{I}ntegration \textbf{T}urbo \textbf{E}ngine (\texttt{unite}: \citealp{hviding_unite_2026}). A local linear continuum is fitted simultaneously over $\pm 7,500$ km s$^{-1}$ about the line. The instrumental line spread function (LSF) is convolved into the model, so the fitted widths are intrinsic and can be compared directly with the NIRES H$\alpha$ width in Table \ref{Table: Components}. We fit four component combinations: a single broad Gaussian, a single exponential, Gaussian+narrow, and exponential+narrow to Pa$\beta$ and compare them using the widely applicable information criterion (WAIC). The WAIC significantly prefers the exponential broad component plus a narrow Gaussian, shown in Figure \ref{Fig: NIRSpec XMM4}, giving FWHM$_{\rm broad} = 965 \pm 56$ km s$^{-1}$. A broad Gaussian is strongly disfavoured as the true profile (Gaussian+narrow, $\Delta$WAIC $=9.88$).

Broad H$\alpha$ from Keck/NIRES gives FWHM $= 965 \pm 55$ km s$^{-1}$ (Table \ref{Table: Components}) against $965 \pm 56$ km s$^{-1}$ for Pa$\beta$. The two measurements are independent, using different instruments, reductions, and fitting codes. The exact agreement is partly coincidental, since the adopted models differ: Pa$\beta$ includes a narrow component that the H$\alpha$ fit does not require (Table \ref{Table: BIC}). Fitting both lines with the same exponential+narrow model gives $1031 \pm 69$ km s$^{-1}$ for H$\alpha$ and $965 \pm 56$ km s$^{-1}$ for Pa$\beta$. The broad Pa$\beta$ flux is $(3.94 \pm 0.13) \times 10^{-17}$ erg s$^{-1}$ cm$^{-2}$, giving an observed broad H$\alpha$/Pa$\beta$ of $10.2 \pm 0.6$ against a Case B expectation of $17.6$.
\section{Discussion}
\label{Sec: Discussion}

\subsection{Confirming LRDs with ground-based selection} \label{Sec: Cosmic Noon LRDs}

Our goal is to determine which of our candidates are consistent with LRDs. Many properties are used to identify LRDs: a V-shaped continuum, broad Balmer emission with non-Gaussian wings, Balmer absorption, elevated Balmer decrements, a compact morphology, and X-ray non-detections \citep{kocevski_rise_2025, degraaff_little_2025}. 

The V-shaped continuum, point-source morphology and broad lines largely occur together \citep{hviding_rubies_2025}, but not all LRDs show all these properties. Samples show that 40--60\% of LRDs show Balmer absorption \citep{davis_oceans_2026, matthee_engine_2026}. In RUBIES, published photometric selections identify only $50$--$80\%$ of the spectroscopic LRDs \citep{hviding_rubies_2025}. No single criterion can therefore decide whether an individual object is, or is not, an LRD. We instead compare our candidates to the JWST-selected population property by property. Variability and the cold dust continuum are beyond these data altogether.

All six candidates show broad H$\alpha$ and Balmer decrements well above the Case B value (Section \ref{Sec: Ionisation}). All six are best described by an exponential broad profile rather than Gaussian (Figure \ref{Fig: Velocity}). Four of the six candidates show more properties consistent with LRDs: XMMLSS-z2LRD1, XMMLSS-z2LRD2, ECOSMOS-z2LRD1 and DEEP23-z2LRD1. These four are consistent with continuum-slope criteria of \cite{kocevski_rise_2025} (Figure \ref{Fig: Beta Slopes}, left) and none is X-ray detected, although only XMMLSS-z2LRD2 and ECOSMOS-z2LRD1 have Chandra coverage (Section \ref{Sec: X-ray}). Three of these four also show Balmer absorption. This 75\% is above the 40--60\% found in larger samples \citep{davis_oceans_2026, matthee_engine_2026}, although it is based on only four objects. The absorption velocities also resemble those in published LRD samples. In \cite{davis_oceans_2026} and \cite{matthee_engine_2026}, the Balmer absorbers have median velocities of $-49$ km s$^{-1}$ and $-60$ km s$^{-1}$, respectively. Our median is consistent at $-37$ km s$^{-1}$.

XMMLSS-z2LRD2 additionally shows broad Pa$\beta$, broad HeI with absorption, and FeII emission. It also shows Calcium triplet absorption that is common in LRDs \citep{greene_calcium_2026}. XMMLSS-AGN-1 does not meet the slope criteria and is the only X-ray detected candidate of the six, at $L_X = 1.5\times10^{44}$ erg s$^{-1}$ (Section \ref{Sec: X-ray}) and $L_{\mathrm{bH}\alpha}/L_X = 0.18$. These properties together make XMMLSS-AGN-1 better described as an ordinary broad-line AGN than as a canonical LRD, although not all LRDs are X-ray undetected \citep{kocevski_rise_2025}. It may also represent a transition phase between the LRD and unobscured AGN populations such as in the X-ray dot \citep{hviding_xray_2026} or \cite{fu_discovery_2025}. XMMLSS-AGN-2 lies just outside the UV slope limit (Section \ref{Sec: Beta Slopes}) and has a strong \textit{Spitzer}/MIPS 24\,$\mu$m detection. This rest-frame mid-infrared emission indicates hot dust, which LRDs typically lack \citep{perez-gonzalez_what_2024, leung_exploring_2025, setton_confirmed_2025}. We therefore do not class XMMLSS-AGN-2 as an LRD.

Chandra observations of XMMLSS-z2LRD2 and ECOSMOS-z2LRD1 show they are X-ray weak by at least factors of $\approx5$ and $\approx7$, respectively, relative to \cite{jin_combined_2012a}, consistent with the X-ray weakness broadly reported for the LRD population \citep{ananna_x-ray_2024, ji_lord_2025, maiolino_jwst_2025}. 


Of the 24 candidates observed, only 10 have a spectroscopic redshift, and 8 of those place H$\alpha$ in our coverage. Four of these eight targets with spectroscopically confirmed H$\alpha$ coverage are LRDs ($50\%$). If the photometric redshifts are taken at face value, 18 targets have H$\alpha$ in coverage, giving 4/18 ($22\%$). Of the six sources in which we detect broad H$\alpha$, four are LRDs ($67\%$). If the photometric redshifts of the remaining targets are correct, those with H$\alpha$ in coverage but no detection must have broad H$\alpha$ fainter than $\approx 3 \times 10^{42}$ erg s$^{-1}$ at $z = 2.1$, or $\approx 6 \times 10^{42}$ erg s$^{-1}$ at $z = 2.9$ (Section \ref{Sec: Results}). These may be fainter LRDs, or they may not be LRDs at all. XMMLSS-LAE-1, for example, shows no broad H$\alpha$ although its spectroscopic redshift places H$\alpha$ in coverage, and it is detected in X-rays at an ordinary AGN luminosity (Section \ref{Sec: X-ray}). Deeper spectroscopy is needed to distinguish between incorrect photometric redshifts, faint broad lines, and misclassification.

At least in DEEP23-z2LRD1 and ECOSMOS-z2LRD1, narrow H$\alpha$, when left free, has a width consistent with [OIII] (Section \ref{Sec: Line Modelling}), pointing to a common origin in a standard narrow-line region (NLR). XMMLSS-z2LRD1 may also host a standard NLR with an additional outflowing ionised gas component.

The principal result is the recovery of four sources with a combination of continuum and line properties resembling \textit{JWST}-selected LRDs. Their presence at cosmic noon shows that the conditions producing these signatures are not confined to the early universe.

\subsection{Separating LRDs from reddened quasars} \label{Sec: Reddened Quasars}
We next examine which properties distinguish our LRDs from reddened quasars selected by similar broadband colors.

\cite{stepney_rest_2026} analyse 63 HRQs at $0.7 < z < 2.7$. These HRQs were also selected to be red, compact objects. Their HRQs are up to 2 orders of magnitude brighter in $K$ band compared to our LRD selected sample. Four of their HRQs meet the LRD continuum slope selections and are described as massive cosmic noon analogues, three of which fall in our redshift range but only two have the correct slopes (Section \ref{Sec: Beta Slopes}). One of these two objects, ULASJ2315+0143, is highly X-ray luminous at $L_X=10^{45.6}$ erg s$^{-1}$ and has a H$\alpha$ FWHM of $4040$ km s$^{-1}$ \citep{stepney_big_2024}. These properties would likely place ULASJ2315+0143 close to the F2M sample in Figure \ref{fig: Lum vs FWHM}. Therefore the continuum shape alone does not separate HRQs from LRDs. More properties that could separate them, such as Balmer absorption, Balmer decrements above Case B, cannot be tested from the \cite{stepney_rest_2026} sample because only the SED photometry of these HRQs is publicly available.

The F2M quasars have much broader H$\alpha$ lines than our LRDs do. Their 9 objects span FWHM $= 3,820$--$6,135$ km s$^{-1}$ with a median of $\approx 4,500$ km s$^{-1}$, against $951-1,428$ km s$^{-1}$ for the four LRDs in Section \ref{Sec: Cosmic Noon LRDs} (Table \ref{Table: Components}). Seven of the nine F2M lines are best described by two Gaussians, broad and narrow, and the other two by a Lorentzian. The 9 F2M quasars with usable measurements have a median Balmer decrement of $6.1$, lower than ours. The F2M sample also shows no signs of Balmer absorption which is common in LRDs. F2M object \texttt{J154938.80+124507.9} incidentally does meet the LRD slope criteria, although only barely. This object only has two optical points (UV $=4$) and so is not well sampled, as is the rest of the F2M sample. Figure \ref{fig: Lum vs FWHM} clearly separates these objects from the population of LRDs.

\subsection{Broad-line widths, black hole masses and accretion rates} \label{Sec: Biases} \label{Sec: BHM} \label{Sec: Eddington Ratio}
We compare our broad H$\alpha$ measurements with other LRD samples and assess what standard AGN calibrations imply for black hole masses and accretion rates.

The systematic uncertainties on $M_{\rm BH}$ are large: $M_{\rm BH}$ assumes the broad line width traces virial motion, and $L_{\rm bol}$ assumes a bolometric correction calibrated on unobscured type 1 AGN, neither of which can be assumed to be valid for this population. The virial assumption also does not hold if the broad line is dominated by scattering \citep{rusakov_little_2026}. In \cite{greene_what_2025}, it is discussed that the H$\alpha$ equivalent width is $\approx 2-5$ times higher than in standard AGN that \cite{greene_estimating_2005} utilise, meaning this bolometric luminosity calibration may overestimate the true $L_{\rm bol}$ (and therefore $M_{\rm BH}$) by as much as $\approx 20$ times. \cite{juodzbalis_direct_2025} report a direct $M_{\rm BH}$ measurement of the LRD QSO1 that may be consistent with the virial estimates; however, this is a single object: both the mass and bolometric luminosity calibrations in LRDs still remain uncertain and object dependent.

We calculate $M_{\rm BH}$ from the H$\alpha$ broad line following \cite{greene_estimating_2005} with Equation \ref{Eq: MBH} below. We note that although we use the standard virial calibration, we do not imply it is correct, given the uncertainties above. Instead, we highlight that the bolometric luminosity has systematic errors far larger.

\begin{equation}
    \label{Eq: MBH}
    \begin{split}
        M_{\rm BH} = (2.0^{+0.4}_{-0.3}) \times 10^6 \left(\frac{L_{\textrm{H}\alpha}}{10^{42}\ \textrm{erg\  s}^{-1}}\right)^{0.55 \pm 0.02} \\ 
        \times \left(\frac{\textrm{FWHM}_{\textrm{H}\alpha}}{10^3\ \textrm{km\ s}^{-1}}\right)^{2.06 \pm 0.06} \ M_{\odot} 
    \end{split}    
\end{equation}

We calculate $L_{\rm bol}$ directly from the broad H$\alpha$ luminosity using the empirical relation of \cite{stern_type_2012}.

\begin{equation}
    \label{Eq: Lbol}
    L_{\rm bol} = 130 \times^{2.4}_{\div 2.4}\ L_{\mathrm{bH}\alpha}
\end{equation}

$L_{\mathrm{bH}\alpha}$ is the luminosity of the broad H$\alpha$ component alone, listed in Table \ref{Table: Components}, and $\times^{2.4}_{\div 2.4}$ denotes the systematic uncertainty of the relation. The uncertainties quoted in Table \ref{Table: Results} propagate the measurement error on $L_{\mathrm{bH}\alpha}$ only.

The Eddington ratio is $\lambda_{\rm Edd} = L_{\rm bol} / L_{\rm Edd}$, where $L_{\rm Edd}$ is the standard Eddington luminosity for ionised hydrogen, $1.26 \times 10^{38}\ (M_{\rm BH}/M_\odot)$ erg s$^{-1}$. Exact values for all six targets are given in Table \ref{Table: Results}.

\begin{table}[ht!]
    \centering
    \caption{Derived properties of the six sources with broad H$\alpha$. Quoted uncertainties are measurement errors only and exclude the $0.38$ dex dispersion of \cite{stern_type_2012}.}
    \setlength{\tabcolsep}{4pt}
    \begin{tabular}{cccc}
        \toprule
        \textbf{Target} & $M_{\mathrm{BH}}$ & $L_{\mathrm{bol}}$ & $\lambda_{\mathrm{Edd}}$ \\
                    & $10^6$ \textbf{$[M_\odot]$} & $10^{45}$ \textbf{[erg/s]} & \\
        \midrule
        XMMLSS-AGN-2      & $74.2 \pm 16$    & $5.25 \pm 0.25$   & $0.56 \pm 0.12$ \\
        XMMLSS-z2LRD1      & $9.40 \pm 2.1$   & $2.62 \pm 0.088$  & $2.21 \pm 0.48$ \\
        XMMLSS-z2LRD2      & $7.66 \pm 1.7$   & $1.70 \pm 0.078$  & $1.77 \pm 0.39$ \\
        XMMLSS-AGN-1     & $46.4 \pm 14$    & $3.51 \pm 0.40$   & $0.60 \pm 0.18$ \\
        ECOSMOS-z2LRD1   & $15.5 \pm 3.6$   & $2.48 \pm 0.11$   & $1.28 \pm 0.29$ \\
        DEEP23-z2LRD1  & $23.9 \pm 6.8$   & $3.10 \pm 0.15$   & $1.03 \pm 0.29$ \\
        \botrule
    \end{tabular}
    \label{Table: Results}
\end{table}

\begin{figure}
    \centering
    \includegraphics[width=\linewidth]{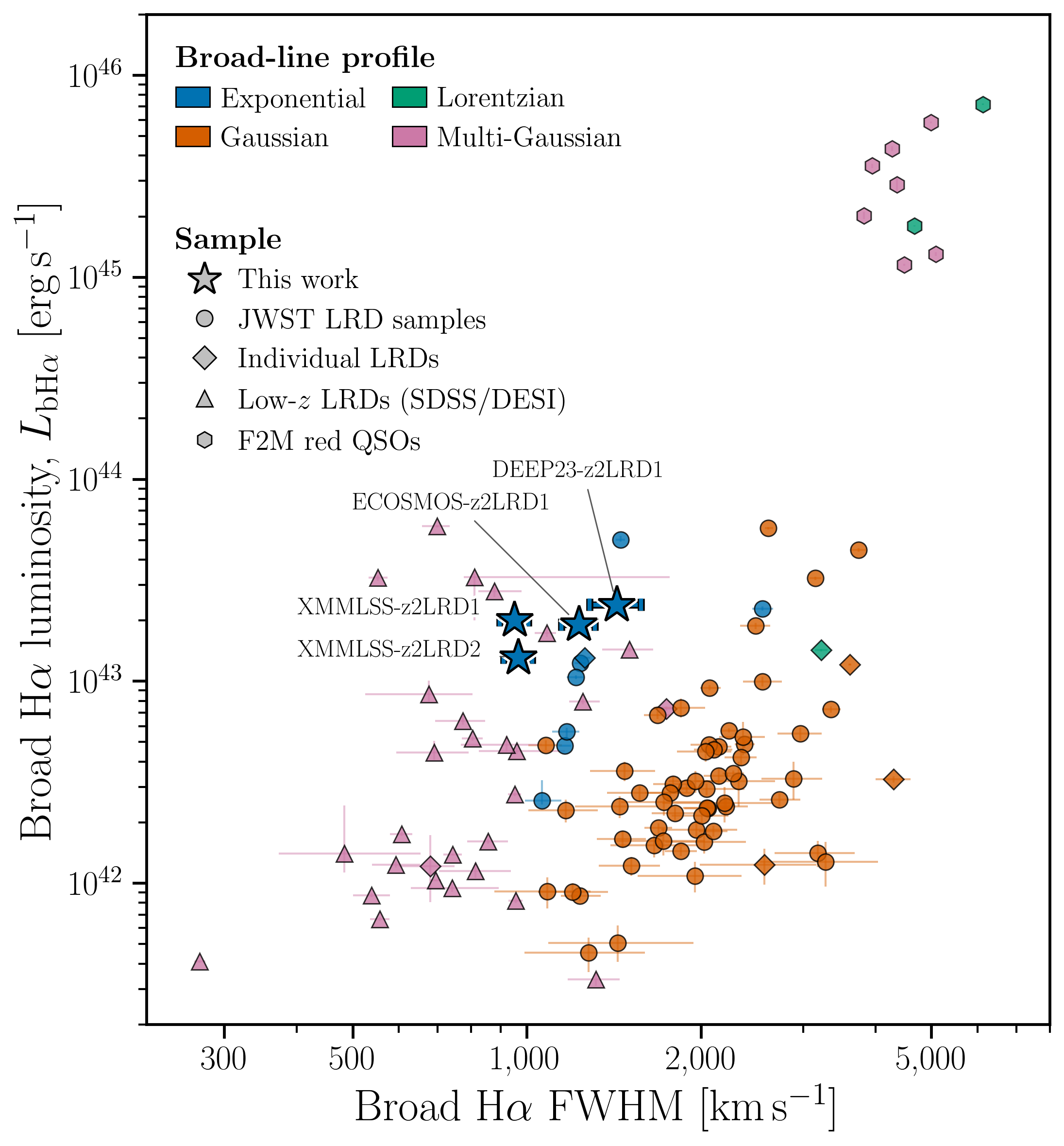}
    \caption{Broad H$\alpha$ luminosity against broad H$\alpha$ FWHM for our four LRDs (stars), LRDs from the literature, and F2M red quasars \citep{glikman_first2mass_2012} (hexagons). Colour gives the profile of the adopted broad component and marker shape the sample.}
    \label{fig: Lum vs FWHM}
\end{figure}

Figure \ref{fig: Lum vs FWHM} compares our broad H$\alpha$ measurements with LRDs from the literature. The comparison sample comprises JWST LRD samples, namely \cite{rusakov_little_2026, taylor_broad-line_2025, matthee_little_2024, zhang_abundant_2026}. We also include low-redshift LRDs from SDSS and DESI \citep{lin_discovery_2026, lin_lrds2_2026}, as well as a variety of individual LRDs \citep{juodzbalis_jades_2024, torralba_black_2026, deugenio_jades_2025, kokorev_silencing_2024, hviding_xray_2026, deugenio_blackthunder_2026}. The F2M red quasars are our own fits to the spectra of \cite{glikman_first2mass_2012} (Section \ref{Sec: Reddened Quasars}). Where only fluxes are published we convert them to luminosities with our cosmology. Our exponential widths of FWHM $\approx 950$--$1430$ km s$^{-1}$ lie among the exponential fits of \cite{rusakov_little_2026}. It is clear the F2M sample is a distinct population set apart from LRDs.

Under the standard AGN calibrations adopted here, our four LRDs have nominal Eddington ratios of $\lambda_{\rm Edd} = 1.0$--$2.2$ (Table \ref{Table: Results}), although the systematic uncertainties prevent a firm determination of their accretion rates. The two LRDs with usable Chandra coverage are X-ray weak relative to their broad H$\alpha$ luminosities (Section \ref{Sec: X-ray}). This weakness could reflect high accretion rates or substantial gas obscuration \citep{madau_xray_2024, maiolino_jwst_2025}; the present observations do not distinguish these possibilities. Dedicated deeper X-ray observations at cosmic noon would provide stronger constraints.

The elevated Balmer decrements resemble those reported in other LRDs \citep{deugenio_irony_2025, inayoshi_extremely_2025, torralba_warm_2025} and indicate similar conditions exist at cosmic noon. Their origin remains uncertain: dense-gas radiative transfer or obscuration may contribute \citep{inayoshi_extremely_2025, madau_little_2026}. ALMA observations of other LRD samples constrain their dust emission, although translating these constraints into optical attenuation depends on the dust geometry \citep{setton_confirmed_2025, casey_upper_2025}. Our observations do not establish which mechanism dominates.

XMMLSS-z2LRD2 provides an additional comparison between hydrogen lines. Its broad H$\alpha$ and Pa$\beta$ widths agree within their uncertainties (Section \ref{Sec: NIRSpec}), although we note this agreement alone does not distinguish scattering from gas kinematics \citep{rusakov_little_2026, scholtz_little_2026}. Under the adopted dust-screen assumptions, the broad H$\alpha$/Pa$\beta$ ratio implies $A_V = 1.1 \pm 0.1$ mag and predicts H$\alpha$/H$\beta \approx 4.0$, below the inferred Balmer decrement. Additional hydrogen-line measurements and multi-epoch spectroscopy could help constrain the origin of the line profiles and ratios.

\section{Conclusion}
\label{Sec: Conclusion}

We present Keck/NIRES, VLT/X-shooter, and Magellan/FIRE spectroscopic follow-up observations of photometrically selected LRD candidates at cosmic noon ($2 < z < 3$), drawn from a parent sample of 266 candidates identified in deep HSC imaging. 

\begin{itemize}[leftmargin=*]
    \item Four candidates: XMMLSS-z2LRD1, XMMLSS-z2LRD2, ECOSMOS-z2LRD1, and DEEP23-z2LRD1 (previously reported by \citealp{ma_counting_2025}) show the majority of the properties that characterise the JWST-selected LRD population: broad H$\alpha$ emission, V-shaped continua, Balmer absorption, elevated Balmer decrements, and X-ray weakness. XMMLSS-AGN-1 shows the fewest of these and is detected in X-rays at a luminosity typical of unobscured AGN. Of the remaining 18 targets, four have Ly$\alpha$ redshifts, which place H$\alpha$ outside our spectral coverage for two. The other 14 have only photometric redshifts, so their H$\alpha$ coverage is uncertain, and they are ambiguous. Those whose photometric redshifts place H$\alpha$ in our coverage must either have incorrect redshifts or broad H$\alpha$ fainter than $L_{\mathrm{H}\alpha} \approx 3 \times 10^{42}$ erg s$^{-1}$. Four of the eight targets with spectroscopically confirmed H$\alpha$ coverage are LRDs ($50\%$). If the photometric redshifts are taken at face value, H$\alpha$ is in coverage for 18 of the 24 observed candidates, giving $22\%$. Of the six sources in which we detect broad H$\alpha$, four are LRDs ($67\%$).

    \item The four LRDs have exponential broad H$\alpha$ with FWHM $= 950$--$1430$ km s$^{-1}$ and $L_{\mathrm{bH}\alpha} = 1.3$--$2.4 \times 10^{43}$ erg s$^{-1}$, among the exponential fits to JWST LRDs (Figure \ref{fig: Lum vs FWHM}). Taken at face value, single-epoch virial calibrations \citep{greene_estimating_2005} give BH masses of 8--$24 \times 10^6\ M_\odot$, and the H$\alpha$-based relation of \cite{stern_type_2012} gives bolometric luminosities of $\approx 2$--$3 \times10^{45}$ erg s$^{-1}$. Both follow from calibrations derived for unobscured type 1 AGN and are model dependent with one to two orders of magnitude uncertainty.

    \item The H$\alpha$ emission line profiles reveal a diversity of physical conditions across the four LRDs. XMMLSS-z2LRD1 shows tentative signs of outflowing gas from a blueshifted Balmer absorber and tentative outflowing ionised gas in [OIII]. Higher resolution spectroscopy is needed to confirm both. Redshifted absorption in ECOSMOS-z2LRD1 instead points toward infalling gas at $46 \pm 14\ \rm km\ s^{-1}$ relative to its narrow line, characteristic of infall \citep{hall_broad_2013, shi_redshifted_2016}. The contrasting kinematics are interpreted as different feedback/accretion states through the same dense medium rather than distinct sight-lines, although viewing angle matters if the geometry contains an open channel.

    \item The four LRDs show elevated Balmer decrements, resembling those reported in other LRD samples. Dense-gas radiative transfer and obscuration remain possible explanations, and the present observations do not establish their relative contributions.
    
    \item X-ray observations with Chandra yield non-detections for the two LRDs with usable coverage, with a stacked limit of $L_X < 2.0\times10^{43}$ erg s$^{-1}$. Using the low-redshift relation between X-ray and broad H$\alpha$ luminosity of \cite{jin_combined_2012a}, they are X-ray weak by at least a factor of $\approx 5$, broadly consistent with the X-ray weakness reported for the LRD population \citep{ananna_x-ray_2024, ji_lord_2025, maiolino_jwst_2025}. This may be consistent with Compton-thick obscuration ($N_H \gtrsim 10^{24}$ cm$^{-2}$), but we cannot rule out near/super-Eddington accretion as an alternative that does not require a Compton-thick column along the line of sight \citep{madau_xray_2024}. By contrast, two candidates without broad H$\alpha$, XMMLSS-LAE-1 and XMMLSS-NOLINE-5, are detected at ordinary AGN X-ray luminosities.
    
\end{itemize}


Our observations extend the characteristic continuum and emission-line signatures of LRDs to cosmic noon. Deeper X-ray observations and additional deep spectroscopy of hydrogen-lines with JWST/MIRI offer promising ways to constrain their obscuration and line-emitting gas.

\section*{acknowledgments}

The data products presented herein were retrieved from the Dawn JWST Archive (DJA). DJA is an initiative of the Cosmic Dawn Center (DAWN), which is funded by the Danish National Research Foundation under grant DNRF140.

We thank the PypeIt development team and the members of the PypeIt Users Slack for their assistance with the data reduction.

This research was funded by the Australian Government through the Australian Research Council (ARC). Associate Professor Ivo Labb\'{e} is the recipient of an ARC Future Fellowship (project number FT220100798). Distinguished Professor Karl Glazebrook is the recipient of an ARC Laureate Fellowship (grant number FL180100060).

\newpage
\bibliography{references}
\bibliographystyle{aasjournalv7}

\appendix
\section{Ly$\alpha$ and tentative H$\alpha$ detections}
\label{Sec: Appendix}

Figure \ref{Fig: Lyman Lines} shows Ly$\alpha$ in five VLT/X-shooter targets. Ly$\alpha$ gives the spectroscopic redshifts in Table \ref{Table: Targets} where no H$\alpha$ is available. This places H$\alpha$ in XMMLSS-LAE-3 and XMMLSS-LAE-4 beyond the wavelength coverage. For XMMLSS-LAE-1 and XMMLSS-LAE-2, H$\alpha$ falls within our coverage but is not detected. Figure \ref{Appendix: Appendix Spectra} shows the [OIII] and H$\alpha$ regions and the photometric SEDs of these four LAEs. DEEP23-NOLINE-2 and XMMLSS-NOLINE-8 may show a tentative low SNR H$\alpha$ line in their Keck/NIRES spectra.

\setcounter{figure}{0}
\renewcommand{\thefigure}{A\arabic{figure}}
\renewcommand{\theHfigure}{A\arabic{figure}}

\begin{figure}[ht!]
    \centering
    \includegraphics[width=\linewidth]{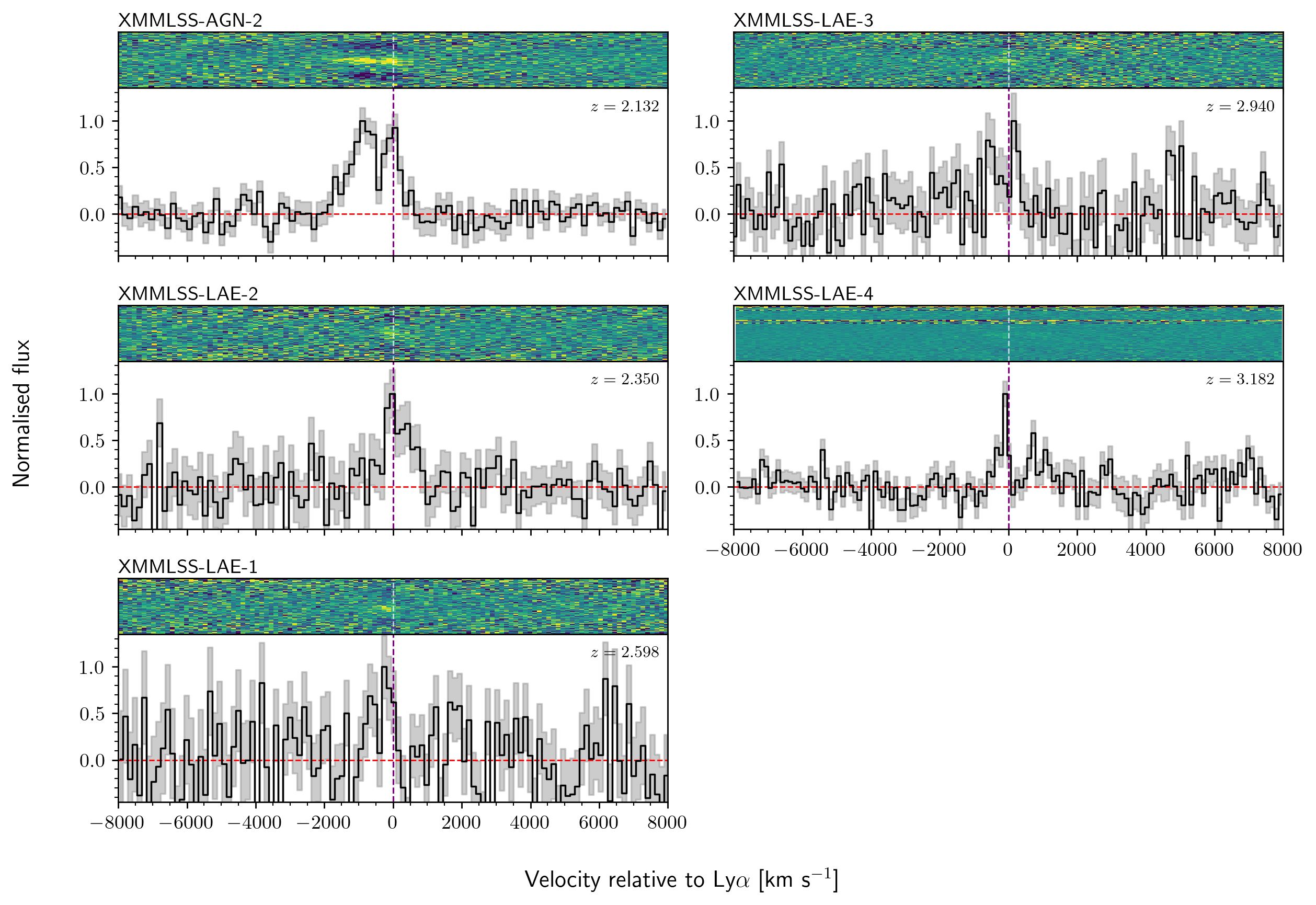}
    \caption{VLT/X-shooter objects with Ly$\alpha$ emission.}
    \label{Fig: Lyman Lines}
\end{figure}

\begin{figure}[ht!]
    \centering
    \includegraphics[width=\linewidth]{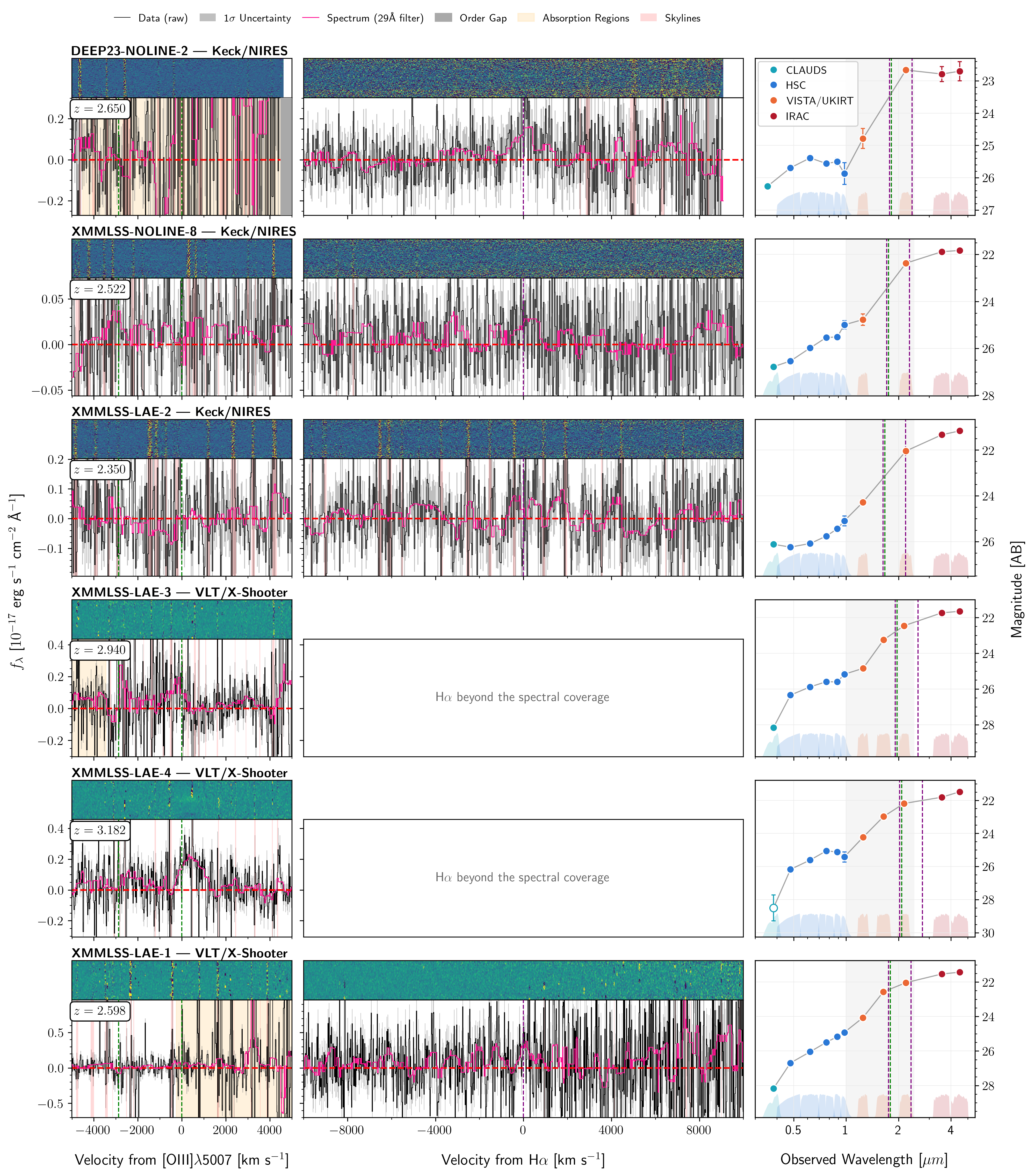}
    \caption{Spectra within $\pm5000$ km s$^{-1}$ of [OIII]$\lambda5007$ and $\pm10000$ km s$^{-1}$ of H$\alpha$, and photometric SEDs, of targets with either a tentative H$\alpha$ or a Ly$\alpha$ line, otherwise shown as in Figure \ref{Fig: Spectra NIRES}.}
    \label{Appendix: Appendix Spectra}
\end{figure}




\end{document}